\documentclass[11pt,a4paper]{article}
\usepackage{jheppub}
\usepackage{bm}
\usepackage{amsmath}
\usepackage{amssymb}

\begin{document}
\newcommand{\llbar}[1]{\bar{\lambda}_{#1}\lambda^{#1}}
\newcommand{\ubbm}{\underline}
\flushbottom
\title{\boldmath Topological Amplitude Computations using the Pure Spinor Formalism}


\author{Nathan Berkovits$^{*}$, Luis Alberto Ypanaqu\'e$^{*}$}

\affiliation{$^{*}$ICTP South American Institute for Fundamental Reserch\\
Instituto de F\'{i}sica Te\'{o}rica, UNESP-Universidade Estadual Paulista\\ R. Dr. Bento T. Ferraz 271, Bl. II, S\~{a}o Paulo 01140-070, SP, Brazil}

\emailAdd{nathan.berkovits@unesp.br, ypanaque.rocha@unesp.br}

\abstract{After constructing a simplified four-dimensional version of the $b$ ghost, topological multiloop amplitudes in type II superstring theory compactified on a six-dimensional orbifold are computed using the non-minimal pure spinor formalism.  These pure spinor amplitude computations preserve manifest $N=2$ $D=4$ supersymmetry and, unlike the analogous
topological multiloop amplitude computations using the hybrid formalism, can be extended to non-topological amplitudes.}

\hfill{}
\maketitle


\section{Introduction}

The connection between certain amplitudes in type II superstring theories compactified on a $\hat{c}=3$ $N=(2,2)$ superconformal field theory (SCFT) and the partition functions of $A$ and $B$ models of the $\hat{c}=3$ topological string has been known since the works of \cite{agnt,bcov}.
The result can be stated as follows: the $g$-loop superstring amplitude of two graviton and $2g-2$ graviphoton string states reproduces at low energies the effective field theory term $F_g R^2 T^{2g-2}$ where $F_g$ is the genus $g$ partition function of a $\hat{c}=3$ topological string and $R$ and $T$ are the self-dual components of the curvature and
graviphoton field-strength. 
The computation was first done for orbifold compactifications
in which the internal variables are free worldsheet fields, and then generalized to arbitrary $\hat c=3$ Calabi-Yau compactifications.

Manifestly target-space supersymmetric formulations of superstrings avoid the necessity of summing over spin structures, simplifying string amplitude computations. Since type II superstrings compactified on a $\hat{c}=3$ $N=(2,2)$ superconformal field theory have $N=2$ spacetime supersymmetry in four dimensions, an appropriate formulation that manifestly preserves this amount of target-space supersymmetry is the hybrid formalism \cite{nbCY}. This hybrid formalism involves Green-Schwarz-like variables for the $D=4$ superspace, RNS-like variables for the internal six-dimensional compactification, and left- and right-moving chiral bosons, $\rho$ and $\tilde{\rho}$. Like the chiral bosons $\phi$ and $\tilde\phi$ coming from bosonization of RNS superconformal ghosts, $\rho$ and $\tilde\rho$ have negative energy and lead to subtleties in multiloop amplitude computations. Although these subtleties can be resolved for topological multiloop amplitudes \cite{bvN4,ktv}, they have up to now prevented the multiloop computation of non-topological amplitudes using the hybrid formalism.

An alternative superstring formalism with manifest target-space supersymmetry is the pure spinor formalism \cite{nbPS} which contains $D=10$ Green-Schwarz-like superspace variables. In addition, it contains bosonic worldsheet variables $\lambda^{\ubbm\alpha}\ (\ubbm\alpha=1,...,16)$, which are spacetime spinors and satisfy the pure spinor constraint $\lambda^{\ubbm\alpha}\gamma^m_{\ubbm\alpha\ubbm\beta}\lambda^{\ubbm\beta}=0$ for $m=1$ to 10. After including non-minimal variables which decouple from the BRST cohomology and constructing a composite $b$ ghost, this pure spinor formalism \cite{nbNMPS} has been used in a flat $D=10$ background to compute multiloop superstring amplitudes up to three loops.
Since the pure spinor formalism includes the 16 $\theta$ variables of $D=10$ superspace, orbifold and Calabi-Yau compactifications of this formalism contain many more worldsheet variables than those of the hybrid formalism which only contains 4 $\theta$ variables. Nevertheless, it will be shown here that a simplified version of the composite $b$ ghost can be constructed for orbifold compactifications if one only requires $D=4$ super-Poincar\'e covariance. This four-dimensional version of the $b$ ghost depends on all 16 $\theta$'s and is equivalent to the usual ten-dimensional $b$ ghost up to a BRST-exact term, but has a simpler form and only depends on two components of the non-minimal variables.
Using this simplified $b$ ghost on an orbifold compactification, it is straightforward to show that the standard pure spinor amplitude prescription correctly computes the $N=2$ $D=4$ supersymmetric version of the $F_g R^2 T^{2g-2}$ topological
term in the Type II effective action. Work is in progress on using this simplified $b$ ghost to also compute non-topological amplitudes in an orbifold compactification.
 
Section \ref{sec:psorbi} of this paper contains a description of the worldsheet variables for an orbifold compactification to four dimensions of the type II non-minimal pure spinor superstring. Section \ref{sec:bghost} discusses the construction of the $b$ ghost both in the original ten-dimensional version and in the simplified four-dimensional version.
Finally, section \ref{sec:topamp} presents the amplitude prescription using the non-minimal pure spinor formalism compactified on an orbifold, and the genus $g$ topological amplitudes are shown to reduce at low energies to the partition function $F_g$ of the internal model times an $N=2$ $D=4$ supersymmetric term containing $R^2 T^{2g-2}$. In section \ref{sec:conclusions}, conclusions are presented and possible future directions of research are proposed. Appendix \ref{app:fourb} shows the BRST equivalence of the ten-dimensional and four-dimensional $b$ ghosts, Appendix \ref{app:twistedzm} explains the zero mode structure of twisted fields, and Appendix \ref{app:twistedpf} reviews the holomorphic factorization of the twisted partition function which appears in the amplitude computation.

\section{Orbifold Compactification of the Type II Pure Spinor Superstring}\label{sec:psorbi}

The pure spinor formalism in a flat background \cite{nbPS} is defined by adding to the usual Green-Schwarz-Siegel $D=10$ superspace worldsheet fields $(x^m,\theta^{\ubbm\alpha},p_{\ubbm\alpha})$ a set of bosonic ghosts $\lambda ^{\ubbm\alpha}$ which, from the target-space point of view, behave as pure spinors of the same chirality as $\theta^{\ubbm\alpha}$. Pure spinors exist in any even dimension, and in the particular case of $D=10$, they are defined as Weyl spinors $\lambda^{\ubbm\alpha}$ satisfying the set of algebraic relations
\begin{equation}
\lambda^{\ubbm\alpha}\gamma^m_{\ubbm\alpha\ubbm\beta}\lambda^{\ubbm\beta}=0,
\end{equation}
where ${\ubbm\alpha}=1,...16$ and $m=1,...,10$. These constraints reduce the number of independent components of the pure spinor from sixteen to eleven.
The pure spinor $\lambda^{\ubbm\alpha}$, whose components have conformal weight $0$, enters into the worldsheet action together with its conjugate momentum $w_{\ubbm\alpha}$ of opposite chirality, which has components of conformal weight $+1$. This variable $w_{\ubbm\alpha}$ also has eleven degrees of freedom because the pure spinor constraint allows for the gauge invariance $\delta w_{\ubbm\alpha}=\Lambda_m(\gamma^m\lambda)_{\ubbm\alpha}$.
The physical spectrum of the superstring is reproduced by the cohomology of the nilpotent charge, $Q=\int dz\lambda^{\ubbm\alpha}d_{\ubbm\alpha}$, which is known as the pure spinor BRST charge, although it is not completely understood how this object appears as a result of BRST gauge-fixing. The field $d_{\ubbm\alpha}$ that appears in the BRST charge is the target space supersymmetric derivative, defined as
\begin{equation}
d_{\ubbm\alpha}=p_{\ubbm\alpha}-{1\over 2}(\gamma^m\theta)_{\ubbm\alpha}\partial x_m-{1\over 8}(\gamma^m\theta)_{\ubbm\alpha}(\theta\gamma_m\partial\theta);
\end{equation}
and the spacetime supersymmetry charge is $q_{\ubbm\alpha} = \int dz j_{\ubbm\alpha} $ where
\begin{equation}
j_{\ubbm\alpha}=p_{\ubbm\alpha}+{1\over 2}(\gamma^m\theta)_{\ubbm\alpha}\partial x_m+{1\over 24}(\gamma^m\theta)_{\ubbm\alpha}(\theta\gamma_m\partial\theta).
\end{equation}

The massless physical states of the open string are given by the ghost number one vertex operator with zero conformal weight, $V=\lambda^{\ubbm\alpha}A_{\ubbm\alpha}$, where $A_{\ubbm\alpha}(x,\theta)$ only depends on the worldsheet zero modes of $x$ and $\theta$. The BRST-closed condition implies that $A_{\ubbm\alpha}$ satisfies the linearized ten-dimensional super-Yang-Mills superfield equations of motion with gauge invariances coming from the BRST exactness of trivial states. In the case of type II superstrings, there are also right-moving worldsheet fields. For type IIB, left-moving spinors $\theta^{\ubbm\alpha}$ and right-moving spinors $\tilde{\theta}^{\ubbm\alpha}$ have the same target-space chirality; while for type IIA, right-moving spinors $\tilde{\theta}_{\ubbm\alpha}$ have chirality opposite to that of $\theta^{\ubbm\alpha}$. The massless closed string vertex operator is a left-right product and has the form $V=\lambda^{\ubbm\alpha}\tilde{\lambda}^{\ubbm\beta}A_{\ubbm\alpha\ubbm\beta}$ \Big($V=\lambda^{\ubbm\alpha}\tilde{\lambda}_{\ubbm\beta}A_{\ubbm\alpha}^{\ \ \ubbm\beta}$\Big) for type IIB (IIA). Equations of motion and gauge invariance come from cohomology requirements on both left- and right-moving sectors by means of BRST charges $Q$ and $\tilde{Q}$, and describe linearized type IIB (IIA) supergravity in ten dimensions.

In the next subsection, this pure spinor construction will be adapted to the  case of type II superstrings compactified on an orbifold to four-dimensional Minkowski spacetime.

\subsection{Orbifold compactification}
To define the string model, it is convenient to perform a split of the worldsheet fields which preserves four-dimensional Lorentz-invariance and the structure of the internal manifold. In the case of Calabi-Yau compactifications, this internal manifold has $SU(3)$ holonomy, so it is convenient to use $U(3)\equiv SU(3)\times U(1)_{\rm int}$ indices for internal directions of the worldsheet fields which come from vectors of the target space. Thus, spacetime coordinates $x^m$, $m=1,...10$ split as $x^\mu$, $\mu=1,...,4$ for flat uncompactified spacetime, while the remaining six internal directions combine  to give complex coordinates $x^I$ and $x_I$ for $I=1$ to 3, transforming in the (anti-)fundamental of $SU(3)$, and carrying internal $U(1)_{\rm int}$ charge $+1\ (-1)$.
 In other words, $x^{I=1}={1\over\sqrt{2}}(x^5+ix^6),\ x_{I=1}={1\over\sqrt{2}}(x^5-ix^6)$, etc. Equivalent notation is $(x^I, x^{\bar{I}})$ where the upper index $\bar{I}$ can be lowered by use of a hermitian metric $g_{I\bar{J}}$, which in the case of orbifold compactifications is just the flat metric, $g_{I\bar{J}}=\delta_{I\bar{J}}$.

It is also necessary to decompose the sixteen component spinors using four-dimensional and internal indices. For example, chiral spinors in ten dimensions $\zeta^{\ubbm\alpha}$ decompose as
\begin{equation}\label{breakchiral}
(\zeta_\alpha,\zeta_{\dot{\alpha}},\zeta_\alpha^I,\zeta_{\dot{\alpha}I}).
\end{equation}
where $(\alpha, \dot\alpha)=1$ to 2 are four-dimensional chiral and anti-chiral indices. 
Similarly, anti-chiral spinors in ten dimensions $\bar{\zeta}_{\ubbm\alpha}$ decompose as
\begin{equation}\label{breakanti}
(\bar{\zeta}_\alpha,\bar{\zeta}_{\dot{\alpha}},\bar{\zeta}_{\alpha I},\bar{\zeta}_{\dot{\alpha}}^I).
\end{equation}
where a bar on a spinor does not necessarily mean it is anti-chiral.

It is convenient to have expressions for different contractions between spinors and gamma matrices in terms of these split components. The relevant formulas are
\begin{equation}\label{scalarcontrac}
\zeta^{\ubbm\alpha}\bar{\zeta}_{\ubbm\alpha}=\zeta^{\alpha}\bar{\zeta}_\alpha-\zeta_{\dot{\alpha}}\bar{\zeta}^{\dot{\alpha}}-\zeta^{\alpha I}\bar{\zeta}_{\alpha I}+\zeta_{\dot{\alpha}I}\bar{\zeta}^{\dot{\alpha}I},
\end{equation}

\begin{equation}\label{4doneform}
\zeta^{\ubbm\alpha}\gamma^\mu_{\ubbm\alpha\ubbm\beta}\xi^{\ubbm\beta}=\zeta^{\alpha}\sigma^\mu_{\alpha\dot{\beta}}\xi^{\dot{\beta}}+\xi^{\alpha}\sigma^\mu_{\alpha\dot{\beta}}\zeta^{\dot{\beta}}-\zeta^{\alpha I}\sigma^\mu_{\alpha\dot{\beta}}\xi^{\dot{\beta}}_{\ I}-\xi^{\alpha I}\sigma^\mu_{\alpha\dot{\beta}}\zeta^{\dot{\beta}}_{\ I},
\end{equation}
\begin{equation}\label{su3up}
\zeta^{\ubbm\alpha}\gamma^I_{\ubbm\alpha\ubbm\beta}\xi^{\ubbm\beta}=\zeta^{\alpha}\xi_\alpha^{\ I}-\zeta^{\alpha I}\xi_\alpha+\varepsilon^{IJK}\zeta_{\dot{\alpha}J}\xi^{\dot{\alpha}}_{\ K},
\end{equation}
\begin{equation}\label{su3down}
\zeta^{\ubbm\alpha}\gamma_{I\ubbm\alpha\ubbm\beta}\xi^{\ubbm\beta}=-2\zeta_{\dot{\alpha} I}\xi^{\dot{\alpha}}+2\zeta_{\dot{\alpha}}\xi^{\dot{\alpha}}_I-2\varepsilon_{IJK}\zeta^{\alpha J}\xi^{\ K}_{\alpha},
\end{equation}

\begin{equation}
\bar{\zeta}_{\ubbm\alpha}\gamma^{\mu\ubbm\alpha\ubbm\beta}\bar{\xi}_{\ubbm\beta}=-\bar{\zeta}^{\alpha}\sigma^\mu_{\alpha\dot{\beta}}\bar{\xi}^{\dot{\beta}}-\bar{\xi}^{\alpha}\sigma^\mu_{\alpha\dot{\beta}}\bar{\zeta}^{\dot{\beta}}+\bar{\zeta}^{\alpha}_{\ I}\sigma^\mu_{\alpha\dot{\beta}}\bar{\xi}^{\dot{\beta}I}+\bar{\xi}^{\alpha}_{\ I}\sigma^\mu_{\alpha\dot{\beta}}\bar{\zeta}^{\dot{\beta}I},
\end{equation}
\begin{equation}
\bar{\zeta}_{\ubbm\alpha}\gamma^{I\ubbm\alpha\ubbm\beta}\bar{\xi}_{\ubbm\beta}=-\bar{\zeta}_{\dot{\alpha}}\bar{\xi}^{\dot{\alpha}I}+\bar{\zeta}_{\dot{\alpha}}^{\ I}\bar{\xi}^{\dot{\alpha}}-\varepsilon^{IJK}\bar{\zeta}^{\alpha}_{\ J}\bar{\xi}_{\alpha K},
\end{equation}
\begin{equation}
\bar{\zeta}_{\ubbm\alpha}\gamma_I^{\ubbm\alpha\ubbm\beta}\bar{\xi}_{\ubbm\beta}=2\bar{\zeta}^\alpha_{\ I}\bar{\xi}_\alpha-2\bar{\zeta}^{\alpha}\bar{\xi}_{\alpha I}+2\varepsilon_{IJK}\bar{\zeta}_{\dot{\alpha}}^{\ J}\bar{\xi}^{\dot{\alpha}K},
\end{equation}
where $\sigma^\mu_{\alpha\dot\alpha}$ are four-dimensional Pauli matrices.

Orbifold compactifications can now be defined as in \cite{dhvf1,dhvf2} where 
the internal manifold of the model is a quotient of the six-torus $T^6$ by a discrete subgroup of its isometry group, $G$. That is, different points of $T^6$, $\{x^I,x_I\}$ and $\{x'^I,x'_I\}$, are identified if they are related by the action of $g\in G$, $x\sim x'=gx$. Fixed points under $G$ become singularities in the quotient, but the string theory is still well-defined.

Physical states of this theory are divided into two sectors. Those inherited from the parent toroidal compactification are said to belong to the untwisted sector of the Hilbert space. Then there are twisted sectors where strings are not closed in the parent $T^6$, but are closed after orbifold identification. In both of these sectors, only $G$-invariant states are allowed and all non-invariant states with respect to $G$ are projected out.

Vertex operators are constructed out of fields which obey periodic or twisted boundary conditions. Insertions of \emph{twisted} vertex operators satisfy \emph{twisted} correlation functions which obey the appropriate monodromies around their insertions; this is similar to the case of spin field insertions on the worldsheet. In the present work concerning topological amplitudes of the compactified superstring, all vertex operators belong to the untwisted sector, so it is not necessary to discuss twisted correlation functions. However, for higher genus amplitudes one should consider nontrivial boundary conditions around each homology cycle of the genus $g$ Riemann surface $\Sigma_g$ on which the amplitude is defined. Each possible choice of boundary conditions is called a twist structure, and the amplitude computations have to include a sum over all twist structures, including the trivial one where boundary conditions of the worldsheet fields are periodic around all $a_j$ and $b_j$ cycles. 
Untwisted worldsheet fields have the usual integer mode expansions and twisted ones have in general non-integer mode expansions.

Bosonic orbifold coordinates $x^I$ and $x_I$ are defined such that
\begin{equation}\label{xtw}
x^I(z+a_j)=e^{2\pi i h_I^{(a_j)}}x^I(z),\hspace{0.5cm}x^I(z+b_j)=e^{2\pi i h_I^{(b_j)}}x^I(z),
\end{equation}
\begin{equation}
x_I(z+a_j)=e^{-2\pi i h_I^{(a_j)}}x_I(z),\hspace{0.5cm}x_I(z+b_j)=e^{-2\pi ih_I^{(b_j)}}x_I(z),
\end{equation}
where $\big\{h_I^{(a_j)},h_I^{(b_j)}\big\}$ specify the orbifold group elements that twist the boundary conditions along the homology cycles of $\Sigma_g$.

Boundary conditions for other worldsheet fields which have tangent indices $I$ are similarly defined by requiring that all conserved currents of the theory are well-defined. For example, twisted boundary conditions for $\theta^I_\alpha$ and $\theta_{\dot{\alpha}I}$ (and corresponding right-moving worldsheet fields) are obtained by requiring the correct amount of supersymmetry to be preserved. This means that components $j_\alpha$ and $j_{\dot{\alpha}}$ of the supersymmetry currents must be well-defined or, in  other words, single-valued along the homology cycles. These currents, together with their right moving analogs, define the $D=4$ $N=2$ supersymmetry of the model and their explicit expressions are
\begin{equation}
j_\alpha=p_\alpha+{1\over 2}\sigma^\mu_{\alpha\dot{\beta}}\theta^{\dot{\beta}}\partial x_\mu+{1\over 2}\theta^I_\alpha\partial x_I+{1\over 24}\sigma^\mu_{\alpha\dot{\beta}}\theta^{\dot{\beta}}(\theta^{\ubbm\alpha}\gamma_{\mu\ubbm\alpha\ubbm\beta}\partial\theta^{\ubbm\beta})+{1\over 24}\theta^I_\alpha(\theta^{\ubbm\alpha}\gamma_{I\ubbm\alpha\ubbm\beta}\partial\theta^{\ubbm\beta}),
\end{equation}
\begin{equation}
j_{\dot{\alpha}}=p_{\dot{\alpha}}+{1\over 2}\theta^\beta\sigma^\mu_{\beta\dot{\alpha}}\partial x_\mu-\theta_{\dot{\alpha}I}\partial x^I+{1\over 24}\theta^\beta\sigma^\mu_{\beta\dot{\alpha}}(\theta^{\ubbm\alpha}\gamma_{\mu\ubbm\alpha\ubbm\beta}\partial\theta^{\ubbm\beta})-{1\over 12}\theta_{\dot{\alpha}I}(\theta^{\ubbm\alpha}\gamma^I_{\ubbm\alpha\ubbm\beta}\partial\theta^{\ubbm\beta}).
\end{equation}
For these currents to be single-valued, one needs to impose 
\begin{equation}
\theta_\alpha^I(z+c_j)=e^{2\pi ih_I^{(c_j)}}\theta_\alpha^I(z),\hspace{1cm}\theta_{\dot{\alpha}I}(z+c_j)=e^{-2\pi ih_I^{(c_j)}}\theta_{\dot{\alpha}I}(z),
\end{equation}
where $c_j$ can be $a_j$ or $b_j$, and $U(1)_{\rm int}$ charge conservation is almost enough to imply
single-valuedness. The only additional requirement comes from the following terms in the supersymmetry currents:
\begin{equation}
j_\alpha=...-{1\over 12}\varepsilon_{IJK}\theta^I_\alpha\theta^{\beta J}\partial\theta^K_\beta,\hspace{0.7cm}j_{\dot{\alpha}}=...-{1\over 12}\varepsilon^{IJK}\theta_{\dot{\alpha}I}\theta_{\dot{\beta}J}\partial\theta^{\dot{\beta}}_K.
\end{equation}
These terms are not single-valued unless $\exp(2\pi i(h_1^{(c_j)}+h_2^{(c_j)}+h_3^{(c_j)}))=1$, which is a well-known constraint, together with the right-moving analog, to obtain $N=2$ $D=4$ spacetime supersymmetry from orbifold compactifications of type II superstrings. In particular, one can choose
\begin{equation}\label{sumzero}
h_1^{(c_j)}+h_2^{(c_j)}+h_3^{(c_j)}=0.
\end{equation}

Notice that the other components of the initial ten-dimensional supersymmetry current, $j_{\alpha I}$ and $j_{\dot{\alpha}}^I$, are not single-valued when all $h_I\neq 0$, so the amount of supersymmetry is reduced to $N=2$ $D=4$. On the other hand, the trivial twist structure where $h_1=h_2=h_3=0$ preserves all 32 supersymmetry currents, giving rise to a subsector with $N=8$ spacetime supersymmetry. Another possible situation is where one of the $h's$ is zero, e.g. $h_1=0$,  and the other two $h$'s, e.g. $h_3=-h_2$, are non-zero. In this case, besides $j_\alpha$ and $j_{\dot{\alpha}}$, the currents $j_{\alpha 1}$ and $j^1_{\dot{\alpha}}$ are also single-valued, so the amount of supersymmetry is only reduced to $N=4$ $D=4$. It will later be shown that for the $R^2 T^{2g-2}$ topological amplitude computations, orbifolds preserving $N=4$ or $N=8$ supersymmetry do not contribute since there are too many fermionic zero modes to be absorbed.

For the remaining worldsheet variables, boundary conditions for the conjugate momentum should be chosen to imply single-valuedness of the worldsheet action, so
\begin{equation}
p_{\dot{\alpha}}^I(z+c_j)=e^{2\pi ih_I^{(c_j)}}p_{\dot{\alpha}}^I(z),\hspace{0.7cm}p_{\alpha I}(z+c_j)=e^{-2\pi ih_I^{(c_j)}}p_{\alpha I}(z).
\end{equation}
Furthermore, the requirement that the minimal pure spinor BRST current $\lambda^{\ubbm\alpha}d_{\ubbm\alpha}$ is single-valued implies that
\begin{equation}
\lambda_\alpha^I(z+c_j)=e^{2\pi ih_I^{(c_j)}}\lambda_\alpha^I(z),\hspace{0.7cm}\lambda_{\dot{\alpha}I}(z+c_j)=e^{-2\pi ih_I^{(c_j)}}\lambda_{\dot{\alpha}I}(z),
\end{equation}
and that its conjugate variables satisfy the boundary conditions
\begin{equation}
w_{\dot{\alpha}}^I(z+c_j)=e^{2\pi ih_I^{(c_j)}}w_{\dot{\alpha}}^I(z),\hspace{0.7cm}w_{\alpha I}(z+c_j)=e^{-2\pi ih_I^{(c_j)}}w_{\alpha I}(z).
\end{equation}

Notice that condition (\ref{sumzero}) is crucial for the pure spinor constraint to be well defined, independently of the requirement of supersymmetry in four-dimensional spacetime. Using (\ref{4doneform}), (\ref{su3up}), and (\ref{su3down}), the pure spinor conditions $\lambda^{\ubbm\alpha}\gamma^m_{\ubbm\alpha\ubbm\beta}\lambda^{\ubbm\beta}=0$ decompose into three equations
\begin{equation}\label{ps1}
\lambda^\alpha\lambda^{\dot{\beta}}-\lambda^{\alpha I}\lambda^{\dot{\beta}}_I=0,
\end{equation}
\begin{equation}\label{ps2}
\lambda^\alpha\lambda_\alpha^I+{1\over 2}\varepsilon^{IJK}\lambda_{\dot{\alpha}J}\lambda^{\dot{\alpha}}_K=0,
\end{equation}
\begin{equation}\label{ps3}
\lambda_{\dot{\alpha}I}\lambda^{\dot{\alpha}}+{1\over 2}\varepsilon_{IJK}\lambda^{\alpha J}\lambda_\alpha^{K}=0,
\end{equation}
and the last two equations only make sense if (\ref{sumzero}) is satisfied.\\

\subsection{Zero modes of twisted fields}

Twisted boundary conditions on a genus $g$ Riemann surface change the zero mode structure of the worldsheet fields. Note that the zero mode of a field having conformal weight zero cannot satisfy twisted boundary conditions because it should be a constant. For the case of compactified spacetime coordinates, $(x^I,x_I)$, their zero modes are constrained to be the fixed points at the orbifold singularities.

Recall that the worldsheet systems in question are fermionic or bosonic field theories with an action of the form,
\begin{equation}
S=\int_{\Sigma_g}(b\bar{\partial}c+\tilde{b}\partial\tilde{c}).
\end{equation}
For untwisted fields, it is well known that the number of zero modes of $b$ and $c$ are related by the Riemann-Roch theorem \cite{abmnv}
\begin{equation}
	n(b)-n(c)=(2\lambda-1)(g-1)
\end{equation}
where $(b,c)$ have conformal weights $(\lambda, 1-\lambda)$.
As explained in Appendix \ref{app:twistedzm}, the same formula can be applied for twisted  systems. So for the twisted case when $(b,c)$ have conformal weights $(1,0)$,  $c$ has no zero modes and $b$ has $g-1$ zero modes which are given by a basis of the so-called $h$-twisted one-differentials $\omega_{h,i}$ for $i=1,...,g-1$ \cite{agnt,bernardtwisted}.

In the twisted sector of the orbifold compactified model, the fermionic variables $\theta^{\ubbm\alpha}$ will split into superspace coordinates $(\theta^\alpha, \theta^{\dot{\alpha}})$ that are always untwisted, and internal coordinates $(\theta_\alpha^I, \theta_{\dot{\alpha}I})$. So there are four untwisted $(1,0)$ systems, $(p_\alpha,\theta^\alpha)$ and $(p_{\dot{\alpha}},\theta^{\dot{\alpha}})$, six $h_I$-twisted $(1,0)$ systems $(p^\alpha_I,\theta^I_\alpha)$, and six $-h_I$-twisted $(1,0)$ systems $(p^{\dot{\alpha}I},\theta_{\dot{\alpha}I})$, where the distinction between
$h_I$-twisted or $-h_I$-twisted depends on the position
of the $I$ index.
 Concerning the zero modes, the untwisted systems contribute four constant zero modes contained in $(\theta^\alpha,\theta^{\dot{\alpha}})$ and $4g$ zero modes contained in $(d_\alpha, d_{\dot{\alpha}})$, the $-h_I$-twisted systems contribute $6g-6$ zero modes contained in $d^{\dot{\alpha}I}$ and no zero modes for $\theta_{\dot{\alpha}I}$, and
 the $h_I$-twisted systems contribute $6g-6$ zero modes contained in $d^\alpha_I$ and no zero modes for $\theta^I_\alpha$.

In the case of constrained variables, the situation is more involved. As in the previous case, the pure spinor field $\lambda^{\ubbm\alpha}$ has four untwisted, six $h_I$-twisted, and six $-h_I$-twisted components; however, only 11 of these 16 components are independent. Recall that if one breaks $SO(10)$ to $U(5)$, the pure spinor decomposes as $(\lambda^+,\lambda_{ab},\lambda^a)$ where $a=1,...,5$ \cite{nbPS,nekrasov1}. The pure spinor space can be described using sixteen patches where each patch corresponds to the subset where a specific component is required to be nonzero. For example, in the patch where $\lambda^+\neq 0$, one can solve the constraints by expressing the five components $\lambda^a$ in terms of $\lambda^+$ and $\lambda_{ab}$ as
\begin{equation}
	\lambda^a={1\over 8\lambda^+}\varepsilon^{abcde}\lambda_{bc}\lambda_{de}.
\end{equation}

If one wants to preserve $D=10$ Lorentz covariance, all 16 patches of pure spinor space must be considered. However, if one only requires $D=4$
Lorentz invariance, it is sufficient to only consider the two patches where $\lambda^{\dot\alpha}\neq 0$, i.e. the patch $\lambda^{\dot 1}\neq 0$ and
the patch $\lambda^{\dot 2}\neq 0$. If $\xi_{\dot \alpha} \lambda^{\dot\alpha} \neq 0$ for some $\xi_{\dot\alpha}$, one can use the pure spinor constraints
(\ref{ps1}) and (\ref{ps3}) to solve for the two components of $\lambda^\alpha$ and for the three components of $(\lambda_{\dot\alpha} \lambda^{\dot\alpha}_I)$ in terms of the other
11 components as  
\begin{equation}
\lambda_\alpha={\lambda^I_\alpha\Big(\xi_{\dot{\beta}}\lambda^{\dot{\beta}}_I\Big)\over\xi_{\dot{\gamma}}\lambda^{\dot{\gamma}}},\quad \lambda_{\dot{\alpha}}\lambda^{\dot{\alpha}}_I={1\over 2}\varepsilon_{IJK}\lambda^{\alpha J}\lambda_\alpha^K.
\end{equation}
So the 11 independent components of the pure spinor can be chosen as
$\lambda_{\dot{\alpha}}, \lambda^{\alpha I}$ and ${\xi}_{\dot{\beta}}\lambda^{\dot{\beta}}_I$. To recover $D=4$ Lorentz covariance, the spinor $\xi_{\dot{\alpha}}$ will be later replaced by the components $\bar{\lambda}_{\dot{\alpha}}$ of the non-minimal pure spinor $\bar{\lambda}_{\ubbm\alpha}$.
Since $\lambda_{\dot\alpha}$ is always untwisted, the zero modes in their two components are the only independent zero modes in the twisted sectors. 


Due to constraints $(\ref{ps1})$, $(\ref{ps2})$, $(\ref{ps3})$, the model has a gauge invariance that allows to set five components of $w_{\ubbm\alpha}$ equal to zero. In the patch where $\xi_{\dot{\alpha}}\lambda^{\dot\alpha}\neq 0$, it is possible to gauge away $w_\alpha$ and $\xi_{\dot{\alpha}}w^{\dot{\alpha}I}$. Therefore, the pure spinor sector consists of 11 pairs of $(1,0)$ chiral bosons: two untwisted, $(w^{\dot{\alpha}},\lambda_{\dot{\alpha}})$, six $h_I$-twisted, $(w_{\alpha I},\lambda^{\alpha I})$, and three $-h_I$-twisted, $(\lambda_{\dot{\alpha}}w^{\dot{\alpha}I},\ (\xi_{\dot\beta}\lambda^{\dot{\beta}})^{-1}\xi_{\dot{\alpha}}\lambda^{\dot{\alpha}}_I)$. So there are $2g$ $w_{\dot{\alpha}}$, $6g-6$ $w_{\alpha I}$ and $3g-3$ $\lambda_{\dot{\alpha}}w^{\dot{\alpha}I}$ zero modes.

Alternatively, one could have preserved $D=4$ Lorentz covariance by only considering patches where $\xi_\alpha\lambda^\alpha\neq 0$ for some $\xi_\alpha$ (and then replacing this $\xi_\alpha$ by the non-minimal spinor $\bar{\lambda}_\alpha$). In this case, the 11 pairs of $(1,0)$ chiral bosons are: two untwisted, $(w^{{\alpha}},\lambda_{{\alpha}})$, six $-h_I$-twisted, $(w^{\dot\alpha I},\lambda_{\dot\alpha I})$, and three $h_I$-twisted, $(\lambda^{{\alpha}}w_{{\alpha}I},\ (\xi^{\beta}\lambda_{{\beta}})^{-1}\xi^{{\alpha}}\lambda_{{\alpha}}^I)$.
And one gets two zero modes from $\lambda_\alpha$, $2g$ zero modes from $w_\alpha$, $6g-6$ zero modes from $w^{\dot{\alpha}I}$, and $3g-3$ zero modes from $\lambda_{\dot\alpha} w^{\dot\alpha I}$.

\section{Structure of the b Ghost}\label{sec:bghost}

\subsection{Non-minimal variables}\label{subsecnm}

Computing multiloop scattering amplitudes using the pure spinor formalism
is most convenient using non-minimal variables which decouple from the
BRST cohomology by means of the usual quartet mechanism \cite{ko}.
The bosonic non-minimal variables added to the minimal formalism include a pure spinor $\bar{\lambda}_{\ubbm\alpha}$ of opposite ten-dimensional chirality with respect to the minimal pure spinor $\lambda^{\ubbm\alpha}$, and the corresponding conjugate field $\bar{w}^{\ubbm\alpha}$, each with eleven independent components. To complete the quartet, one introduces fermionic variables  $r_{\ubbm\alpha}$ obeying $\bar{\lambda}_{\ubbm\alpha}\gamma^{\ubbm\alpha\ubbm\beta}_m r_{\ubbm\beta}=0$, that has eleven independent components because of this constraint, and its conjugate field $s^{\ubbm\alpha}$, which also has eleven components due to the gauge invariance generated by the constraint. These new degrees of freedom decouple from the cohomology after modifying the minimal BRST charge to
\begin{equation}
Q=\int dz \Big(\lambda^{\ubbm\alpha}d_{\ubbm\alpha}+\bar{w}^{\ubbm\alpha}r_{\ubbm\alpha}\Big).
\end{equation}
Similar considerations work for the right-moving sector of type II theories.

Using the non-minimal variables, a composite $b$-ghost can be constructed
satisfying $\{Q, b\}=T$, where $T=-{1\over 2}\partial x^m\partial x_m-p_{\ubbm\alpha}\partial\theta^{\ubbm\alpha}+w_{\ubbm\alpha}\partial\lambda^{\ubbm\alpha}+\bar{w}^{\ubbm\alpha}\partial\bar{\lambda}_{\ubbm\alpha}-s^{\ubbm\alpha}\partial r_{\ubbm\alpha}$ is the non-minimal stress-energy tensor of the formalism.
The construction uses a chain of operators $(T, G^{\ubbm\alpha},H^{[\ubbm\alpha\ubbm\beta]},K^{[\ubbm\alpha\ubbm\beta\ubbm\gamma]},L^{[\ubbm\alpha\ubbm\beta\ubbm\gamma\ubbm\delta]})$ satisfying the following relations
\begin{equation}\label{descent1}
\{Q,G^{\ubbm\alpha}\}=\lambda^{\ubbm\alpha}T, \hspace{0.5cm} [Q,H^{[\ubbm\alpha\ubbm\beta]}]=\lambda^{[\ubbm\alpha}G^{\ubbm\beta]},\hspace{0.5cm}\{Q,K^{[\ubbm\alpha\ubbm\beta\ubbm\gamma]}\}=\lambda^{[\ubbm\alpha}H^{\ubbm\beta\ubbm\gamma]},
\end{equation}
\begin{equation}\label{descent2}
[Q,L^{[{\ubbm\alpha\ubbm\beta\ubbm\gamma\ubbm\delta}]}]=\lambda^{[\ubbm\alpha}K^{\ubbm\beta\ubbm\gamma\ubbm\delta]}, \hspace{1cm}\lambda^{[\ubbm\alpha}L^{\ubbm\beta\ubbm\gamma\ubbm\delta\ubbm\kappa]}=0.
\end{equation}
One then defines
\begin{equation}\label{bghostnm}
b={\bar{\lambda}_{\ubbm\alpha}G^{\ubbm\alpha}\over \bar{\lambda}\lambda}-{\bar{\lambda}_{\ubbm\alpha}r_{\ubbm\beta}H^{[\ubbm\alpha\ubbm\beta]}\over(\bar{\lambda}\lambda)^2}-{\bar{\lambda}_{\ubbm\alpha}r_{\ubbm\beta}r_{\ubbm\gamma}K^{[\ubbm\alpha\ubbm\beta\ubbm\gamma]}\over(\bar{\lambda}\lambda)^3}+{\bar{\lambda}_{\ubbm\alpha}r_{\ubbm\beta}r_{\ubbm\gamma}r_{\ubbm\delta}L^{[\ubbm\alpha\ubbm\beta\ubbm\gamma\ubbm\delta]}\over(\bar{\lambda}\lambda)^4} +s^{\ubbm\alpha}\partial\bar{\lambda}_{\ubbm\alpha},
\end{equation}
where $\bar{\lambda}\lambda\equiv\bar{\lambda}_{\ubbm\alpha}\lambda^{\ubbm\alpha}$, and 
\begin{equation}\label{chain}
G^{\ubbm\alpha} = {1\over 2} \Pi^m (\gamma_m d)^{\ubbm\alpha} - {1\over 4}N_{mn} (\gamma^{mn} \partial\theta)^{\ubbm\alpha} - {1\over 4} J \partial\theta^{\ubbm\alpha},
\end{equation}
$$H^{[\ubbm\alpha\ubbm\beta]} = {{1}\over{192}} (\gamma^{mnk})^{\ubbm\alpha\ubbm\beta} (d\gamma_{mnk} d + 24 N_{mn} \Pi_k),$$
$$K^{[\ubbm\alpha\ubbm\beta\ubbm\gamma]} = -{{1}\over{96}} (\gamma_m d)^{\left[\ubbm\alpha\right.} (\gamma^{mnk})^{\left.\ubbm\beta\ubbm\gamma\right]} N_{nk},$$
$$
L^{[\ubbm\alpha\ubbm\beta\ubbm\gamma\ubbm\delta]} = -{{1}\over{128}} {{1}\over{4!}} (\gamma_{mnp})^{\left[\ubbm\alpha\ubbm\beta\right.} (\gamma^{pqr})^{\left.\ubbm\gamma\ubbm\delta\right]} N^{mn} N_{qr}.$$ with $\Pi^m = \partial x^m -\theta^{\ubbm\alpha} \gamma^m_{\ubbm\alpha\ubbm\beta} \partial \theta^{\ubbm\beta}$, $N_{mn} = {1\over 2} \lambda \gamma_{mn} w$, and $J=w_{\ubbm\alpha}\lambda^{\ubbm{\alpha}}$.

For orbifold compactifications, boundary conditions for internal non-minimal variables in twisted sectors are implied by single-valuedness of this composite $b$ ghost and BRST charge and are
\begin{equation}
\bar{\lambda}_{\dot{\alpha}}^I(z+c_j)=e^{2\pi ih_I^{(c_j)}}\bar{\lambda}_{\dot{\alpha}}^I(z),\hspace{0.7cm}\bar{\lambda}_{\alpha I}(z+c_j)=e^{-2\pi ih_I^{(c_j)}}\bar{\lambda}_{\alpha I}(z),
\end{equation}
\begin{equation}
\bar{w}_\alpha^I(z+c_j)=e^{2\pi ih_I^{(c_j)}}\bar{w}_\alpha^I(z),\hspace{0.7cm}\bar{w}_{\dot{\alpha}I}(z+c_j)=e^{-2\pi ih_I^{(c_j)}}\bar{w}_{\dot{\alpha}I}(z),
\end{equation}
\begin{equation}
r_{\dot{\alpha}}^I(z+c_j)=e^{2\pi ih_I^{(c_j)}}r_{\dot{\alpha}}^I(z),\hspace{0.7cm}r_{\alpha I}(z+c_j)=e^{-2\pi ih_I^{(c_j)}}r_{\alpha I}(z),
\end{equation}
\begin{equation}
s_\alpha^I(z+c_j)=e^{2\pi ih_I^{(c_j)}}s_\alpha^I(z),\hspace{0.7cm}s_{\dot{\alpha}I}(z+c_j)=e^{-2\pi ih_I^{(c_j)}}s_{\dot{\alpha}I}(z).
\end{equation}

When $\llbar{\dot{\alpha}}\neq 0$, one can use the same arguments as
for the pure spinor $\lambda^{\ubbm\alpha}$ and solve all components of
$\bar\lambda_{\ubbm\alpha}$  in terms of $\bar{\lambda}_{\dot{\alpha}}$, $\bar{\lambda}_{\alpha I}$ and $\lambda_{\dot{\alpha}}\bar{\lambda}^{\dot{\alpha}I}$, and the independent bosonic and fermionic
systems are $(\bar{w}^{\dot{\alpha}},\bar{\lambda}_{\dot{\alpha}})$, $(\bar{w}^{\alpha I},\bar{\lambda}_{\alpha I})$, $((\llbar{\dot{\beta}})^{-1}\bar{\lambda}_{\dot{\alpha}}\bar{w}^{\dot{\alpha}}_I,\ \lambda_{\dot{\alpha}}\bar{\lambda}^{\dot{\alpha}I})$ and 
$(s_{\dot{\alpha}},r^{\dot{\alpha}})$, $(s^{\alpha I},r_{\alpha I})$, $((\llbar{\dot{\beta}})^{-1}\bar{\lambda}_{\dot{\alpha}}s^{\dot{\alpha}}_I,\ \lambda_{\dot{\alpha}}r^{\dot{\alpha}I})$.
Also, when $\llbar{\dot{\alpha}}\neq 0$ in a twisted sector, the bosonic zero modes include two from $\bar{\lambda}_{\dot{\alpha}}$, $2g$ from $\bar{w}_{\dot{\alpha}}$, $6g-6$ from $\bar{w}^{\alpha I}$, and $3g-3$ from $(\llbar{\dot{\beta}})^{-1}\bar{\lambda}_{\dot{\alpha}}\bar{w}^{\dot{\alpha}}_I$. And the fermionic zero modes include  two from $r_{\dot{\alpha}}$, $2g$ from $s_{\dot{\alpha}}$, $6g-6$ from $s^{\alpha I}$, and $3g-3$ from $(\llbar{\dot{\beta}})^{-1}\bar{\lambda}_{\dot{\alpha}}s^{\dot{\alpha}}_I$.\\ 

Because of the inverse factors of $(\lambda^{\ubbm\alpha}\bar\lambda_{\ubbm\alpha})$ in the composite $b$ ghost of (\ref{bghostnm}), there can be divergences in the correlation functions
when $\lambda^{\ubbm\alpha}\to 0$. Since the path integral measure factor $\int d^{11}\lambda d^{11}\bar\lambda$ converges like $(\lambda^{\ubbm\alpha}\bar\lambda_{\ubbm\alpha})^{11}$, the poles in the $b$ ghost can cause divergences
if the poles accumulate to order  $(\lambda^{\ubbm\alpha}\bar\lambda_{\ubbm\alpha})^{-11}$ or worse. The restriction to poles of lower order than  $(\lambda^{\ubbm\alpha}\bar\lambda_{\ubbm\alpha})^{-11}$ can also be understood from BRST
cohomology arguments since $\{Q, \xi\}=1$ where
$$\xi = {{\bar\lambda_{\ubbm\alpha}\theta^{\ubbm\alpha}}\over{\lambda^{\ubbm\alpha}\bar\lambda_{\ubbm\alpha} + r_{\ubbm\alpha}\theta^{\ubbm\alpha}}}
= 
{{\bar\lambda_{\ubbm\alpha}\theta^{\ubbm\alpha}}\over{\lambda^{\ubbm\alpha}\bar\lambda_{\ubbm\alpha} }} + ... + 
{{(\bar\lambda_{\ubbm\alpha}\theta^{\ubbm\alpha})(r_{\ubbm\alpha}\theta^{\ubbm\alpha})^{10}}\over{(\lambda^{\ubbm\alpha}\bar\lambda_{\ubbm\alpha})^{11} }}.$$
So allowing states with $(\lambda^{\ubbm\alpha}\bar\lambda_{\ubbm\alpha})^{-11}$
dependence like $\xi$ into the Hilbert space would trivialize the BRST cohomology
since $Q(\xi V) = V$ whenever $QV=0$.

As will now be shown, there is an alternative four-dimensional $b^{(a)}$ version of the $b$ ghost satisfying $\{Q, b^{(a)}\}=T$ if one restricts to the patches where $\lambda^{\dot\alpha}\neq 0 $, or equivalently, $\bar\lambda_{\dot\alpha}\lambda^{\dot{\alpha}}\neq 0 $. This four-dimensional version satisifes
$b^{(a)} = b + Q \Lambda$ where $\Lambda$ is well-defined when $\bar\lambda_{\dot\alpha}\lambda^{\dot{\alpha}}\neq 0 $, and
\begin{equation}\label{bfour}
b^{(a)}=s^{\ubbm\alpha}\partial\bar{\lambda}_{\ubbm\alpha}+{{\bar{\lambda}_{\dot\alpha}G^{\dot\alpha}}\over {\bar\lambda_{\dot\alpha}\lambda^{\dot\alpha}}}-{{\bar{\lambda}_{\dot\alpha}r_{\dot\beta}H^{[\dot\alpha\dot\beta]}}\over{(\bar{\lambda}_{\dot\alpha}\lambda^{\dot\alpha})^2}}
\end{equation}
$$=s^{\ubbm\alpha}\partial\bar{\lambda}_{\ubbm\alpha}-{{\bar{\lambda}_{\dot\alpha}}\over {\bar\lambda_{\dot\alpha}\lambda^{\dot\alpha}}} \Big({1\over 2} \Pi_I d^{\dot\alpha I} +{1\over 2} \Pi^\mu \tilde\sigma_\mu^{\dot\alpha\beta} d_\beta \Big)-w_{\dot{\alpha}}\partial\theta^{\dot{\alpha}}-w^\alpha_I\partial\theta^I_\alpha$$
$$-{1\over\llbar{\dot{\alpha}}}\Big(\lambda_{\dot{\alpha}}w^{\dot{\alpha}I}\bar{\lambda}_{\dot{\beta}}\partial\theta^{\dot{\beta}}_I-\bar{\lambda}_{\dot{\beta}}\lambda^{\dot{\beta}}_Iw^I_{\dot{\alpha}}\partial\theta^{\dot{\alpha}}-w_{\dot{\alpha}}\lambda^{\dot{\alpha}}\bar{\lambda}_{\dot{\beta}}\partial\theta^{\dot{\beta}}\Big)+{\bar{\lambda}_{\dot{\alpha}}r^{\dot{\alpha}}d^\alpha d_\alpha\over 4(\llbar{\dot{\alpha}})^2}-{\bar{\lambda}_{\dot{\alpha}}r^{\dot{\alpha}}\lambda_{\dot{\beta}}w^{\dot{\beta}I}\Pi_I\over(\llbar{\dot{\alpha}})^2}.$$

One can similarly define a four-dimensional version $b^{(c)}$ satisfying 
$\{Q, b^{(c)}\}=T$ if one restricts to the patches where $\lambda^{\alpha}\neq 0 $ as
\begin{equation}\label{bfourtwo}
b^{(c)}=s^{\ubbm\alpha}\partial\bar{\lambda}_{\ubbm\alpha}+{{\bar{\lambda}_{\alpha}G^{\alpha}}\over {\bar\lambda_{\alpha}\lambda^{\alpha}}}-{{\bar{\lambda}_{\alpha}r_{\beta}H^{[\alpha\beta]}}\over{(\bar{\lambda}_\alpha \lambda^\alpha)^2}}
\end{equation}
$$=s^{\ubbm\alpha}\partial\bar{\lambda}_{\ubbm\alpha}+{{\bar{\lambda}_{\alpha}}\over {\bar\lambda_{\alpha}\lambda^{\alpha}}} \Big({1\over 2} \Pi^I d^{\alpha}_I +{1\over 2} \Pi^\mu \tilde\sigma_\mu^{\dot\alpha\alpha} d_{\dot\alpha} \Big)+w^{\alpha}\partial\theta_{\alpha}+w_{\dot{\alpha}}^{I}\partial\theta^{\dot{\alpha}}_I$$
$$-{1\over\llbar{\alpha}}\Big(\lambda^{\alpha}w_{\alpha I}\bar{\lambda}^{\beta}\partial\theta^{ I}_\beta-\bar{\lambda}^{\beta}\lambda^{I}_\beta w^{\alpha}_I\partial\theta_{\alpha}-w^{\alpha}\lambda_{\alpha}\bar{\lambda}^{\beta}\partial\theta_{\beta}\Big)+{\bar{\lambda}^{\alpha}r_{\alpha}d_{\dot\alpha} d^{\dot\alpha}\over 4(\llbar{\alpha})^2}-{\bar{\lambda}^{\alpha}r_{\alpha}\lambda^{\beta}w_{\beta I}\Pi^I\over(\llbar{\alpha})^2}.$$

Although $b^{(a)}$ is simpler than $b$ of  (\ref{bghostnm}), the restriction to patches $\lambda^{\dot\alpha}\neq 0 $ allows one to define 
$$\xi^{(a)} = {{\bar\lambda_{\dot\alpha}\theta^{\dot\alpha}}\over{\lambda^{\dot\alpha}\bar\lambda_{\dot\alpha} + r_{\dot\alpha}\theta^{\dot\alpha}}}
= 
{{\bar\lambda_{\dot\alpha}\theta^{\dot\alpha}}\over{\lambda^{\dot\alpha}\bar\lambda_{\dot\alpha} }}  -
{{(\bar\lambda_{\dot\alpha}\theta^{\dot\alpha})(r_{\dot\alpha}\theta^{\dot\alpha})}\over{(\lambda^{\dot\alpha}\bar\lambda_{\dot\alpha})^{2} }},$$
which satisfies $\{Q, \xi^{(a)}\}=1$ and only diverges like $(\lambda^{\dot\alpha}\bar\lambda_{\dot\alpha})^{-2}$. Furthermore, in the twisted sector,
$\lambda^{\ubbm\alpha}$ has only two independent zero modes so the path
integral $\int d^2 \lambda d^2 \bar\lambda$ converges only like
$(\lambda^{\dot\alpha}\bar\lambda_{\dot\alpha})^{2}$. So unlike the original
non-minimal formalism where the Hilbert space allows states with poles less
divergent than $(\lambda^{\ubbm\alpha}\bar\lambda_{\ubbm\alpha})^{-11}$, the non-minimal formalism restricted to patches $\lambda^{\dot\alpha}\neq 0 $ only allows
states with poles less divergent than $ (\lambda^{\dot\alpha}\bar\lambda_{\dot\alpha})^{-2}$.

A simple way to obtain $b^{(a)}$ is to start with the ten-dimensional $b$ ghost of
(\ref{bghostnm}) and rescale the non-minimal variables as
\begin{equation}
\bar\lambda_{\alpha I}\to c \bar\lambda_{\alpha I},\quad\quad
\lambda^{\dot\alpha} \bar\lambda_{\dot\alpha}^I \to 
c \lambda^{\dot\alpha} \bar\lambda_{\dot\alpha}^I,
\end{equation}
\begin{equation}
r_{\alpha I}\to c r_{\alpha I},\quad\quad
\lambda^{\dot\alpha} r_{\dot\alpha}^I \to 
c \lambda^{\dot\alpha} r_{\dot\alpha}^I ,
\end{equation}
whereas $\bar\lambda_{\dot\alpha}$ and $r_{\dot\alpha}$ are kept invariant.
From the pure spinor constraints, this implies that 
\begin{equation}
\bar\lambda_{\alpha }\to c^2 \bar\lambda_{\alpha },\quad
\bar\lambda^{\dot\alpha} \bar\lambda_{\dot\alpha}^I \to 
c^2 \bar\lambda^{\dot\alpha} \bar\lambda_{\dot\alpha}^I,
\end{equation}
\begin{equation}
r_{\alpha }\to c^2 r_{\alpha },\quad
\bar\lambda^{\dot\alpha} r_{\dot\alpha}^I \to 
c^2 \bar\lambda^{\dot\alpha} r_{\dot\alpha}^I.
\end{equation}
To leave the worldsheet action and BRST operator invariant, the conjugate non-minimal variables must be rescaled as
\begin{equation}
\bar w^{\alpha I}\to c^{-1} \bar w^{\alpha I},\quad
\bar\lambda_{\dot\alpha} \bar w^{\dot\alpha}_I \to 
c^{-1} \bar\lambda_{\dot\alpha} \bar w^{\dot\alpha}_I 
\end{equation}
\begin{equation}
s^{\alpha I}\to c^{-1} s^{\alpha I},\quad
\bar\lambda_{\dot\alpha} s^{\dot\alpha}_I \to 
c^{-1} \bar\lambda_{\dot\alpha} s^{\dot\alpha}_I
\end{equation}
whereas $\bar w^{\dot\alpha}$ and $s^{\dot\alpha}$ are kept invariant.

Under this rescaling of the non-minimal variables, one can easily verify that in the limit where $c\to 0$, the $b$ ghost
of (\ref{bghostnm}) goes to $b^{(a)}$ of (\ref{bfour}). Since the non-minimal variables do not appear in the BRST-invariant vertex operators, the only other effect of this rescaling is to change the definition of $\Lambda = \bar\lambda_{\ubbm\alpha}
\theta^{\ubbm\alpha} + ...$ which appears in the regulator
 ${\mathcal{N}}= \exp (Q\Lambda)$.  But since ${\mathcal{N}} = 1 + Q(\Lambda +{1\over 2} \Lambda Q\Lambda + ...)$, changing the definition of $\Lambda$ is a BRST-trivial
 operation and does not affect scattering amplitudes. For this reason, one is free to take the limit $c\to 0$ when computing the scattering amplitude which replaces the $b$ ghost
 of (\ref{bghostnm}) with $b^{(a)}$. Of course, there is a different rescaling of the non-minimal variables which
instead replaces $b$ with $b^{(c)}$.

\subsection{Construction of four-dimensional $b$ ghost}

Suppose one restricts to a patch in pure spinor space where either $\llbar{\alpha}\neq 0$ or $\llbar{\dot{\alpha}}\neq 0$. Since $\{Q_0, G^\alpha\}=\lambda^\alpha T_0$ and $\{Q_0, G^{\dot{\alpha}}\}=\lambda^{\dot{\alpha}}T_0$ where $Q_0=\int \lambda^{\ubbm\alpha} d_{\ubbm\alpha}$ and $T_0 =-{1\over 2}\partial x^m\partial x_m-p_{\ubbm\alpha}\partial\theta^{\ubbm\alpha}+w_{\ubbm\alpha}\partial\lambda^{\ubbm\alpha}$ are the minimal BRST charge and stress-tensor,
one can choose on these patches the first term of the four-dimensional $b$ ghost as either
\begin{equation}
b^{(c)}={\bar{\lambda}_\alpha G^\alpha\over\llbar{\beta}}+...,\hspace{1cm}{\rm or}\hspace{1cm}b^{(a)}={\bar{\lambda}_{\dot{\alpha}}G^{\dot{\alpha}}\over\bar{\lambda}_{\dot{\beta}}\lambda^{\dot{\beta}}}+....
\end{equation}
Applying the non-minimal BRST charge to the first term in $b^{(a)}$, one gets
\begin{equation}\label{GT}
\left\{Q_0+\int \bar{w}^{\ubbm\alpha}r_{\ubbm\alpha}, {\bar{\lambda}_{\dot{\alpha}}G^{\dot{\alpha}}\over\bar{\lambda}_{\dot{\gamma}}\lambda^{\dot{\gamma}}}\right\}=T_0-{r_{\dot{\alpha}}G^{\dot{\alpha}}\over\bar{\lambda}_{\dot{\gamma}}\lambda^{\dot{\gamma}}}+{\bar{\lambda}_{\dot{\alpha}}G^{\dot{\alpha}}r_{\dot{\beta}}\lambda^{\dot{\beta}}\over(\bar{\lambda}_{\dot{\gamma}}\lambda^{\dot{\gamma}})^2}.
\end{equation}

To cancel the second and third term in the right hand side of the last equation one proceeds as in the ten dimensional case, picking other operators in the chain. Notice that there is no need for components other than $H^{[\dot{\alpha}\dot{\beta}]}$, which has actually only one independent component since dotted indices take just two values. Using $\Big[Q, H^{[\dot{\alpha}\dot{\beta}]}\Big]=\lambda^{[\dot{\alpha}}G^{\dot{\beta}]}$,

\begin{equation}\label{HG}
\left\{Q_0+\int \bar{w}^{\ubbm\alpha}r_{\ubbm\alpha},- {\bar{\lambda}_{\dot{\alpha}}r_{\dot{\beta}}H^{[\dot{\alpha}\dot{\beta}]}\over(\bar{\lambda}_{\dot{\delta}}\lambda^{\dot{\delta}})^2}\right\} ={r_{\dot{\alpha}}G^{\dot{\alpha}}\over\bar{\lambda}_{\dot{\delta}}\lambda^{\dot{\delta}}}-{\bar{\lambda}_{\dot{\alpha}}G^{\dot{\alpha}}r_{\dot{\beta}}\lambda^{\dot{\beta}}\over(\bar{\lambda}_{\dot{\delta}}\lambda^{\dot{\delta}})^2}+{r_{\dot{\alpha}}r_{\dot{\beta}}H^{[\dot{\alpha}\dot{\beta}]}\over (\bar{\lambda}_{\dot{\delta}}\lambda^{\dot{\delta}})^2}-{2r_{\dot{\gamma}}\lambda^{\dot{\gamma}}\bar{\lambda}_{\dot{\alpha}}r_{\dot{\beta}}H^{[\dot{\alpha}\dot{\beta}]}\over (\bar{\lambda}_{\dot{\delta}}\lambda^{\dot{\delta}})^3}.
\end{equation}
The last two terms in (\ref{HG}) cancel each other because of the identity $r_{\dot{\alpha}}r_{\dot{\beta}}={1\over 2}\varepsilon_{\dot{\alpha}\dot{\beta}}r_{\dot{\gamma}}r^{\dot{\gamma}}$, and the first two cancel those appearing in (\ref{GT}), so we can define the $b$-ghost to be
\begin{equation}
b^{(a)} = {\bar{\lambda}_{\dot{\alpha}}G^{\dot{\alpha}}\over\bar{\lambda}_{\dot{\alpha}}\lambda^{\dot{\alpha}}}-{\bar{\lambda}_{\dot{\alpha}}r_{\dot{\beta}}H^{[\dot{\alpha}\dot{\beta}]}\over(\bar{\lambda}_{\dot{\alpha}}\lambda^{\dot{\alpha}})^2}+s^{\ubbm\alpha} \partial \bar\lambda_{\ubbm\alpha}
\end{equation}
where the last term is necessary to get the non-minimal terms in $T = T_0 +
\bar{w}^{\ubbm\alpha}\partial\bar{\lambda}_{\ubbm\alpha}-s^{\ubbm\alpha}\partial r_{\ubbm\alpha}$.
Similarly, one obtains the other $b$-ghost
\begin{equation}
b^{(c)}={\bar{\lambda}_\alpha G^\alpha\over\llbar{\alpha}}-{\bar{\lambda}_\alpha r_\beta H^{[\alpha\beta]}\over(\llbar{\alpha})^2}+s^{\ubbm\alpha} \partial \bar\lambda_{\ubbm\alpha}.
\end{equation}

Finally, it will be shown in Appendix \ref{app:fourb} that in the patch where $\llbar{\dot{\alpha}}\neq 0$ ($\llbar{\alpha}\neq 0$), the usual $D=10$ pure spinor $b$ ghost of (\ref{bghostnm}) is equivalent to 
$b^{(a)}$ ($b^{(c)}$) up to BRST trivial terms. The formula relating $b$ and $b^{(c)}$ is
\begin{equation}
b={{\bar\lambda_\alpha G^\alpha}\over{\llbar{\alpha}}}-{\bar{\lambda}_\alpha r_\beta H^{[\alpha\beta]}\over(\llbar{\alpha})^2} +s^{\ubbm\alpha} \partial \bar\lambda_{\ubbm\alpha}
\end{equation}
\begin{equation*}
+Q\left({\bar{\lambda}_\alpha\bar{\lambda}_{\beta'}H^{[\alpha,\beta']}\over(\llbar{\alpha})(\llbar{\ubbm\alpha})}+{\bar{\lambda}_\alpha\bar{\lambda}_{\beta'}r_{\ubbm\gamma}K^{[\alpha\beta'\ubbm\gamma]}\over\llbar{\alpha}(\llbar{\ubbm\alpha})^2}+{\bar{\lambda}_\alpha\bar{\lambda}_{\beta'}r_\gamma K^{[\alpha\beta'\gamma]}\over(\llbar{\alpha})^2\llbar{\ubbm\alpha}}\right.
\end{equation*}
\begin{equation*}
\left.-{\bar{\lambda}_\alpha\bar{\lambda}_{\beta'}r_{\ubbm\gamma}r_{\delta'}L^{[\alpha\beta'\ubbm\gamma\delta']}+\bar{\lambda}_\alpha\bar{\lambda}_{\beta'}r_{\gamma'}r_\delta L^{[\alpha\beta'\gamma'\delta]}\over\llbar{\alpha}(\llbar{\ubbm\alpha})^3}-{\bar{\lambda}_\alpha\bar{\lambda}_{\beta'}r_{\gamma'}r_\delta L^{[\alpha\beta'\gamma'\delta]}\over(\llbar{\alpha})^2(\llbar{\ubbm\alpha})^2}\right)
\end{equation*}
where 
the convention used for spinor indices is the following: $\alpha$ denotes chiral spinors in four dimensions, and $\alpha'$ denotes any of the other components in the ten-dimensional quantity, i.e. $\alpha'=(\dot{\alpha},\alpha I, \dot{\alpha}I)$.
So $b=b^{(c)}+Q\Lambda^{(c)}$ in the patch where $\llbar{\alpha}\neq 0$, and one can similarly show that $b=b^{(a)}+Q\Lambda^{(a)}$ in the patch where $\llbar{\dot{\alpha}}\neq 0$.

\section{Amplitude Computation}\label{sec:topamp}

As in a flat $D=10$ background, the $g$-loop $N$-point closed superstring amplitude prescription using the non-minimal pure spinor formalism in an orbifold compactification is 
\begin{equation}\label{pres}
{\mathcal{A}} =\int_{{\mathcal{M}}_g}\left\langle\left|{\mathcal{N}}\prod_{i=1}^{3g-3}b(\mu_i)\right|^2 \prod^{N}_{j=1}U_j\right\rangle,
\end{equation}
where $U_j$ for $j=1,...N$ are vertex operators for the desired amplitude, $b(\mu_i)=\int_\Sigma\mu_i b$ with 
$\mu_i$ for $i=1$ to $3g-3$ a basis of Beltrami differentials on the Riemann surface (and a single $\mu$ for genus one), and ${\mathcal{N}}$ is an appropriately chosen BRST-invariant regulator of the form ${\mathcal{N}}= \exp (Q\Lambda)$ which is inserted anywhere on the Riemann surface and resolves the divergences coming from integration over the non-compact bosonic zero modes. For example, one can define
\begin{equation}
{\mathcal{N}}=\exp\bigg(-\bar{\lambda}_{\ubbm\alpha}\lambda^{\ubbm\alpha}-r_{\ubbm\alpha}\theta^{\ubbm\alpha}-\Big[{1\over 2}N_{mn}\bar{N}^{mn }+J\bar{J}+{1\over 4}S_{mn}d\gamma^{mn}\lambda+S\lambda^{\ubbm\alpha}d_{\ubbm\alpha}\Big]\bigg)
\end{equation}
where
\begin{equation}
N_{mn}={1\over 2}w\gamma_{mn}\lambda,\hspace{1cm}\bar{N}_{mn}={1\over 2}(\bar{w}\gamma_{mn}\bar{\lambda}-s\gamma_{mn}r),\hspace{1cm}J=w_{\ubbm\alpha}\lambda^{\ubbm\alpha},
\end{equation}
\begin{equation}
\bar{J}=\bar{w}^{\ubbm\alpha}\bar{\lambda}_{\ubbm\alpha}-s^{\ubbm\alpha}r_{\ubbm\alpha},\hspace{1cm}S_{mn}={1\over 2}s\gamma_{mn}\bar{\lambda},\hspace{1cm}S=s^{\ubbm\alpha}\bar{\lambda}_{\ubbm\alpha}.
\end{equation}

One would normally define the $b$ ghost as  
 (\ref{bghostnm}), and the amplitude prescription of (\ref{pres}) is well-defined as long as the poles
 from the $b$ ghosts do not accumulate to poles of order $(\lambda^{\ubbm\alpha} \bar\lambda_{\ubbm\alpha})^{-11}$
 when $\lambda^{\ubbm\alpha}\to 0$. Since the charge $\int (\bar w^{\ubbm\alpha} \bar \lambda_{\ubbm\alpha} - s^{\ubbm\alpha} r_{\ubbm\alpha})$  commutes with the $b$ ghost, 
each factor of $(\lambda_{\ubbm\alpha}\bar\lambda^{\ubbm\alpha})^{-1}$ is accompanied by a factor of $r_{\ubbm\alpha}$ and the dangerous terms
are when the $b$ ghosts contribute 11 factors of $r_{\ubbm\alpha}$.
As explained in subsection \ref{subsecnm}, this can be understood either from the
 path integral measure factor which converges like $(\lambda^{\ubbm\alpha} \bar\lambda_{\ubbm\alpha})^{11}$ when
 $\lambda^{\ubbm\alpha} \bar\lambda_{\ubbm\alpha}\rightarrow 0$, or from BRST cohomology arguments. 
 
 But for special
 choices of the external states in the vertex operators $U_j$, one can instead use either the four-dimensional version of
 the $b$ ghost defined as $b^{(a)}$ in (\ref{bfour}), or as $b^{(c)}$ in  (\ref{bfourtwo}). For external states corresponding to anti-self-dual topological amplitudes, one can use the $b^{(a)}$ version since the poles from these $b^{(a)}$ ghosts will not accumulate to poles of order $(\lambda^{\dot\alpha} \bar\lambda_{\dot\alpha})^{-2}$. And for external states corresponding to self-dual topological amplitudes, one can use the $b^{(c)}$ version since the poles from these $b^{(c)}$ ghosts will not accumulate to poles of order $(\lambda^{\alpha} \bar\lambda_{\alpha})^{-2}$. The dangerous terms using these four-dimensional $b$ ghosts appear when the $b^{(a)}$ ghosts contribute $r_{\dot \alpha} r^{\dot\alpha}$  or when the $b^{(c)}$ ghosts contribute $r_\alpha r^\alpha$, and it will be shown from zero-mode counting that these dangerous terms cannot contribute to the topological amplitudes.

\subsection{Universal sector of the spectrum}

The spectrum of type II superstrings compactified on an orbifold to four dimensions consists of a gravitational supermultiplet which contains the graviton and the graviphoton, a universal hypermultiplet which contains the dilaton and the axion and two Ramond-Ramond scalars, and sets of hypermultiplets and vector supermultiplets which, in the case of Calabi-Yau compactifications, are related to the cohomology of the internal manifold \cite{bs4D}. 

To introduce notation, the open superstring vertex operator on an orbifold compactification will be described first. Assuming independence of the internal coordinates, the BRST condition $QV=0$ implies the linearized superfield equations of motion for a four-dimensional $N=1$ super-Maxwell multiplet plus three $N=1$ chiral multiplets \cite{clv1}. Furthermore, since one projects out non-invariant states under the orbifold group, the three chiral multiplets are removed and one is left only with the $N=1$ $D=4$ super-Maxwell multiplet.

For the uncompactified open superstring, the BRST-invariant vertex operator for the on-shell $D=10$ super-Maxwell multiplet is
\begin{equation}\label{vov}
V = \int dz (\partial\theta^{\ubbm\alpha} A_{\ubbm\alpha}(x, \theta) + \Pi^m A_m (x, \theta)+ d_{\ubbm\alpha} W^{\ubbm\alpha} (x, \theta)+ N_{mn} F^{mn}(x, \theta))
\end{equation}
where $(A_{\ubbm\alpha}, A_m)$ are the on-shell spinor and vector superpotentials and $(W^{\ubbm\alpha}, F^{mn})$ are the on-shell spinor and tensor superfields whose lowest
components are the photino and the photon field strength. To obtain the BRST-invariant vertex operator for the on-shell $D=4$ super-Maxwell multiplet from (\ref{vov}), one simply sets
to zero all fields in (\ref{vov}) except for those in the $N=1$ $D=4$ super-Maxwell multiplet. The resulting vertex operator will depend on all 16 $\theta$'s, but will be independent
of the zero modes of the internal coordinates $(x^I, x_I)$ for $I=1$ to 3. 

For the closed superstring in an orbifold compactification, the relevant $N=2$ $D=4$ multiplets are the gravitational multiplet containing the graviton and graviphoton, and the universal hypermultiplet containing the dilaton, axion and complex Ramond-Ramond scalar. The closed string vertex operator for these multiplets is obtained from the left-right product
of two open superstring vertex operators and has the form
\begin{equation}\label{vcv}
V = \int d^2 z  (\partial\theta^{\ubbm\alpha} \bar\partial \tilde\theta^{\ubbm\beta} A_{\ubbm\alpha \ubbm\beta}(x, \theta, \tilde\theta) + ...)
\end{equation}
where $A_{\ubbm\alpha \ubbm\beta}$ is the left-right product of $A_{\ubbm\alpha}$ and $A_{\ubbm\beta}$ and $...$ involves similar left-right products of the other superfields.

As will be shown below, the only terms in the closed superstring vertex operator of (\ref{vcv}) which will contribute to the topological amplitudes are the terms
\begin{equation}\label{vcw}
\int d^2z\ \Big[d^\alpha\tilde{d}^\beta P_{\alpha\beta}+d^{\dot{\alpha}}\tilde{d}^{\dot{\beta}}P_{\dot{\alpha}\dot{\beta}}+d^\alpha\tilde{d}^{\dot{\beta}}Q_{\alpha\dot{\beta}}+d^{\dot{\alpha}}\tilde{d}^{\beta}Q_{\dot{\alpha}\beta}\Big]
\end{equation}
where $(P_{\alpha\beta}, P_{\dot\alpha\dot\beta})$ are superfields whose lowest components are the graviphoton anti-self-dual and self-dual field strengths, and $(Q_{\alpha\dot\beta}, Q_{\dot\alpha\beta})$ are superfields whose lowest components are derivatives of the complex Ramond-Ramond scalar. These $N=2$ $D=4$ superfields can be
understood as the left-right product of chiral and anti-chiral photino $N=1$ $D=4$ superfields, i.e. $P_{\alpha\beta}$ is obtained from the left-right product of $W_\alpha$ with $W_\beta$, $P_{\dot\alpha\dot\beta}$ is obtained from the left-right product of $W_{\dot\alpha}$ with $W_{\dot\beta}$, $Q_{\alpha\dot\beta}$ is obtained from the left-right product of $W_\alpha$ with $W_{\dot\beta}$, and $Q_{\dot\alpha\beta}$ is obtained from the left-right product of $W_{\dot\alpha}$ with $W_\beta$.
Of course, the complete integrated vertex operator will contain additional terms to those of
(\ref{vcw}), but it will be argued that only the terms in (\ref{vcw}) will contribute to the topological amplitudes. This is similar to the situation in the hybrid formalism \cite{bvN4,ktv,bs4D}, but the difference here is that the vertex operators in the pure spinor formalism
depend on all 16 $\theta$ and 16 $\tilde\theta$ variables.

\subsection{Type IIB multiloop scattering of anti-self-dual graviphotons}
 
The first amplitude which will be computed is the $g$-loop type IIB superstring scattering of $2g-2$ anti-self-dual graviphotons and 2 anti-self-dual gravitons which contributes to the $R^2 T^{2g-2}$ term in the low-energy effective action. The computation of the case $g=1$ is slightly different and will be considered later. This multiloop amplitude will be computed using the $b^{(a)}$ ghost of (\ref{bfour}), and because of the integral over the $d_{\ubbm\alpha}$ fermionic zero modes, the only term in the closed string vertex operators
$U_i$ which contributes to this amplitude is the term $\int d^2z d_\alpha \tilde{d}_\beta P^{\alpha\beta}(x^\mu,\theta^{\ubbm\alpha},\tilde{\theta}^{\ubbm\alpha})$ and the only term in the $b^{(a)}$ ghosts which contributes
is $$ b^{(a)} = -{\bar{\lambda}_{\dot{\alpha}}\Pi_I d^{\dot{\alpha}I}\over\llbar{\dot{\alpha}}}.$$

To see this, focus on the zero modes of the left-moving fermionic variables and note that the vertex operators $U_i$ can only contribute the zero modes of the 2 components 
$d_\alpha$ whereas the ghosts $b^{(a)}$ can only contribute the zero modes of the 3 components 
$\bar\lambda_{\dot \alpha} d^{\dot\alpha I}$ and the 2 components $d_\alpha$. The zero modes of the other
11 components of $d_{\ubbm\alpha}$ (i.e. $d_{\dot\alpha}$, $d_{\alpha I}$ and $\lambda_{\dot\alpha} d^{\dot\alpha I}$) must all come from the regulator ${\mathcal{N}}$. Since there are $3g-3$ $b^{(a)}$ ghosts,
the $3g-3$ zero modes of $\bar\lambda_{\dot \alpha} d^{\dot\alpha I}$ must all come from the $b^{(a)}$
ghosts and the $2g$ zero modes of $d_\alpha$ must all come from the $U_i$ vertex operators. Furthermore, only the
twisted sectors of the worldsheet variables contribute to this amplitude since, in the untwisted sector, 
$\bar\lambda_{\dot \alpha} d^{\dot\alpha I}$ would have $3g$ zero modes which cannot be obtained from the
$3g-3$ $b^{(a)}$ ghosts. For the same reason, orbifold sectors which preserve $N=4$ $D=4$ supersymmetry cannot contribute to this amplitude since, in this case, one of the
three $x^I$'s in (\ref{xtw}) would be untwisted and $\bar\lambda_{\dot\alpha} d^{\dot\alpha I}$ would have $3g-2$ zero modes.

In the twisted sector assuming that $\bar\lambda_{\dot\alpha}\lambda^{\dot\alpha}\neq 0$, 
the only fermionic zero modes of $r_{\ubbm\alpha}$ are the two components $r_{\dot\alpha}$ and the
only fermionic zero modes of $\theta^{\ubbm\alpha}$ are the 4 components $\theta^\alpha$ and $\theta^{\dot\alpha}$. In addition to providing the fermionic zero modes of 11 components of $d_{\ubbm\alpha}$, the regulator
${\mathcal{N}}$ also provides the fermionic zero modes of the 11 components of $s^{\ubbm\alpha}$, and the fermionic zero modes of the 2 components of $r_{\dot \alpha}$ and $\theta^{\dot\alpha}$. The remaining zero modes of the two components of $\theta^\alpha$ must come from the vertex operators $U_i$. 

Integration over $d_{\alpha}$ and $\tilde{d}_{\alpha}$ zero modes produces index contractions between the $P^{\alpha\beta}$ superfields, giving the expression 
$$(P_{\alpha\beta}P^{\alpha\beta})^g [{\rm det}(\omega_i(z_j))]^2 [{\rm det}(\bar{\omega}_i(\bar{z}_j))]^2,$$
 where $[{\rm det}(\omega_i(z_j))]^2$ denotes the sum of terms of the form ${\rm det}(\omega_i(z_j)){\rm det}(\omega_i(z_{j'}))$ with $z_j$ denoting $g$ of the $2g$ positions of vertex operators and $z_{j'}$ denoting the other $g$ positions.
Using the relation  
\begin{equation}
	\prod_{i=1}^g\int d^2 z_i\ |{\rm det}(\omega_j(z_k))|^2={\rm det}({\rm Im}\tau),
\end{equation}
and integrating over the zero modes of $(x^\mu, \theta^\alpha, \tilde\theta^\alpha, p_\alpha, \tilde p_\alpha)$, the vertex operators therefore contribute in the low energy limit
$$({\rm det}({\rm Im}\tau))^2 \int d^4 x \int d^2\theta \int d^2\tilde{\theta}(P_{\alpha\beta}(x,\theta, \tilde\theta) P^{\alpha\beta}(x,\theta, \tilde\theta))^g .$$
 
To separate the $\bar\lambda_{\dot\alpha} d^{\dot\alpha I}$ zero modes appearing in the $b^{(a)}$ ghosts from the 
$d^{\dot\alpha I}$ zero modes appearing in the regulator ${\mathcal{N}}$, it is convenient to define a change
of basis for the worldsheet fields as
\begin{equation}\label{change1}
\psi^I=(\bar{\lambda}_{\dot{\beta}}\lambda^{\dot{\beta}})^{-1}\bar{\lambda}_{\dot{\alpha}}p^{\dot{\alpha}I},\hspace{1cm}\chi^I=\lambda_{\dot{\alpha}}p^{\dot{\alpha}I},
\end{equation}
\begin{equation}\label{change2}
\psi_I=\lambda_{\dot{\alpha}}\theta^{\dot{\alpha}}_I,\hspace{1cm}\chi_I=(\bar{\lambda}_{\dot{\beta}}\lambda^{\dot{\beta}})^{-1}\bar{\lambda}_{\dot{\alpha}}\theta^{\dot{\alpha}}_I,
\end{equation}
which is invertible when $\llbar{\dot{\alpha}}\neq 0$ and whose jacobian in the path integral is $1$. The resulting
worldsheet action in terms of these variables is
\begin{equation}
	\int d^2z\ \Big[-\psi^I\bar{\partial}\psi_I+\chi^I\bar{\partial}\chi_I\Big]+S'
\end{equation}
where
\begin{equation}
	S'=\int d^2z\ \Bigg[\Bigg(-\lambda_{\dot{\alpha}}\psi^I\chi_I+{\bar{\lambda}_{\dot{\alpha}}\over\llbar{\dot{\beta}}}\big(\psi^I\psi_I+\chi^I\chi_I\big)\Bigg)\bar{\partial}\lambda^{\dot{\alpha}}+{\bar{\lambda}^{\dot{\alpha}}\chi^I\psi_I\over(\llbar{\dot{\beta}})^2}\bar{\partial}\bar{\lambda}_{\dot{\alpha}}\Bigg]
\end{equation}
and $S'$ can be canceled by shifting $w_{\dot{\alpha}}$ and $\bar{w}^{\dot{\alpha}}$ as
\begin{equation}\label{shift}
	w_{\dot{\alpha}}\longrightarrow w_{\dot{\alpha}}-\lambda_{\dot{\alpha}}\psi^I\chi_I+{\bar{\lambda}_{\dot{\alpha}}\over\llbar{\dot{\beta}}}\big(\psi^I\psi_I+\chi^I\chi_I\big),
\end{equation}
\begin{equation}
	\bar{w}^{\dot{\alpha}}\longrightarrow\bar{w}^{\dot{\alpha}}+{\bar{\lambda}^{\dot{\alpha}}\chi^I\psi_I\over(\llbar{\dot{\beta}})^2},
\end{equation}
in the pure spinor action $\displaystyle\int d^2z\ \Big(w_{\dot{\alpha}}\bar{\partial}\lambda^{\dot{\alpha}}+\bar{w}^{\dot{\alpha}}\bar{\partial}\bar{\lambda}_{\dot{\alpha}}\Big)$.

So after integration over the zero modes of the worldsheet variables, the multiloop amplitude reduces to
\begin{equation}\label{amptopo}
{\mathcal{A}}_g = \int_{{\mathcal{M}}_g} ({\rm det}({\rm Im}\tau))^2  \int d\psi_0^I\int d\tilde\psi_0^I\left\langle\left|\prod_{i=1}^{3g-3}G^-_{int}(\mu_i)\right|^2\right\rangle' \int d^4 x\int d^2\theta\int d^2\tilde{\theta}(P_{\alpha\beta}P^{\alpha\beta})^{g}
\end{equation}
where $\int d^4 x\int d^2\theta\int d^2\tilde{\theta}(P_{\alpha\beta}P^{\alpha\beta})^{g}$ is
the $N=2$ $D=4$ supersymmetric expression containing the term $\int d^4 x R^2 T^{2g-2}$,
$\int d\psi_0^I\int d\tilde\psi_0^I$ denotes integration over the zero modes of $\psi^I$ and $\tilde \psi^I$, $\langle...\rangle'$ means path integration over non-zero modes of all remaining worldsheet fields as well as the lattice sum coming from the $(x^I, x_I)$ path integral, and the only term contributing from
$b^{(a)}$ is
$$G^-_{int}\equiv  - \psi^I \partial x_I.$$ 
Note that the explicit expression for $\Pi_I$ in the $b^{(a)}$ ghost is
 \begin{equation}
\Pi_I=\partial x_I-\theta_{\dot{\alpha}I}\partial\theta^{\dot{\alpha}}-\theta_{\dot{\alpha}}\partial\theta^{\dot{\alpha}}_I-\varepsilon_{IJK}\theta^{\alpha J}\partial\theta^K_\alpha,
\end{equation}
but since all $d_{\ubbm\alpha}$'s must contribute with zero modes, the only term in $\Pi_I$ which
can contribute is $\partial x_I$. For the same reason, the $\theta$ and $\tilde\theta$ dependence of the anti-self-dual graviphoton superfield in the amplitude reduces to $P^{\alpha\beta}(x^\mu,\theta^\alpha,\tilde{\theta}^\alpha)$. Note also that (\ref{amptopo}) holds up to a proportionality factor that comes from integration over pure spinor zero modes. This proportionality factor will not be considered in this work but can be computed using the methods of \cite{gom}.

\subsection*{Non-zero mode integration}

Recall that for a fermionic $(1,0)$ chiral system $(b,c)$ on a Riemann surface of genus $g$, the result for the path integral is
\begin{equation}\label{untwisted}
\int {\mathcal{D}}b{\mathcal{D}}c\ b(z_1)...b(z_g)c(y)e^{-S[b,c]}=Z_1{\rm det}(\omega_i(z_j))
\end{equation}
where $\{\omega_i,i=1,...,g\}$ is a basis for holomorphic $1$-differentials on the genus $g$ Riemann surface $\Sigma_g$, and $Z_1$ is discussed in Appendix \ref{app:twistedpf} and denotes the genus $g$ chiral partition function coming from non-zero mode integration. For bosonic $(1,0)$ chiral systems, the result of the path integral is the inverse $[Z_1{\rm det}(\omega_i(z_j))]^{-1}$.

For the case of worldsheet fields defined within a given non-trivial twist structure $h_I$, the path integral is instead
\begin{equation}
\int{\mathcal{D}}b{\mathcal{D}}c\ b(z_1)...b(z_{g-1})e^{-S[b,c]}=Z_{1,h_I}{\rm det}(\omega_{h_I,i}(z_j))
\end{equation}
where $\{\omega_{h_I,i},i=1,...,g-1\}$ is a basis of $h_I$-twisted holomorphic $1$-differentials and $Z_{1,h_I}$ is the partition function coming from non-zero mode integration for twisted fields. Analogously, the path integral of the bosonic $h_I$-twisted $(1,0)$ system is $[Z_{1,h_I}{\rm det}(\omega_{h_I,i}(z_j))]^{-1}$. As reviewed in Appendix \ref{app:twistedpf}, these formulas can be derived using the general bosonization formula of \cite{abmnv} for the non-chiral theory and performing chiral factorization as in \cite{vvchiral}.

The fermionic variables $(d_{\ubbm\alpha},\theta^{\ubbm\alpha})$ involve four untwisted $(1,0)$ systems $(d_\alpha,\theta^\alpha)$ and $(d_{\dot{\alpha}},\theta^{\dot{\alpha}})$, two copies of the three $h_I$-twisted ($I=1,2,3$) $(1,0)$ systems, $(d_{\dot{\alpha}}^I,\theta^{\dot{\alpha}}_I)$, and two copies of the three $-h_I$-twisted $(1,0)$ systems, $(d_{\alpha I},\theta^{\alpha I})$. Integration over the non-zero modes of $(d_{\ubbm\alpha},\theta^{\ubbm\alpha})$ therefore gives
\begin{equation}
(Z_1)^4\prod_{I=1}^3\Big[(Z_{1,h_I})^2(Z_{1,-h_I})^2\Big].
\end{equation}
So after including the contribution from the right moving sector $(\tilde{d}_{\ubbm\alpha},\tilde{\theta}^{\ubbm\alpha})$, one gets
\begin{equation}
|Z_1|^8\prod_{I=1}^3\Big[|Z_{1,h_I}|^{4}|Z_{1,-h_I}|^{4}\Big].
\end{equation}

Since the bosonic pure spinor variables $(w_{\ubbm\alpha}, \lambda^{\ubbm\alpha})$ are equivalent to eleven bosonic $(1,0)$ chiral systems, nine of them twisted by $h_I$ or $-h_I$ depending on the position of the index $I$ in the corresponding conformal weight one field, their contribution to the amplitude after including the left and right-moving sectors is
\begin{equation}
|Z_1|^{-4}\prod^3_{I=1}\Big[|Z_{1,h_I}|^{-2}|Z_{1,-h_I}|^{-4}\Big].
\end{equation}

Finally, the contribution from the non-zero modes of $x^\mu$ is
$|Z_1|^{-4}  ({\rm det}({\rm Im}\tau))^{-2}$, and the contribution from the bosonic non-minimal variables
$(\bar w^{\ubbm\alpha}, \bar \lambda_{\ubbm\alpha})$ cancels the contribution from the fermionic
non-minimal variables $(s^{\ubbm\alpha}, r_{\ubbm\alpha})$.

Including these contributions from the non-zero modes in (\ref{amptopo}), one therefore obtains
\begin{equation}
{\mathcal{A}}_g = \int_{{\mathcal{M}}_g}\prod_{I=1}^3|Z_{1,h_I}|^2\int d\psi_0^I\int d\tilde\psi_0^I\left\langle\left|\prod_{i=1}^{3g-3}G^-_{int}(\mu_i)\right|^2\right\rangle \int d^4 x \int d^2\theta \int d^2\tilde{\theta}(P_{\alpha\beta}P^{\alpha\beta})^g.
\end{equation}
Finally, one can replace $\prod_{I=1}^3|Z_{1,h_I}|^2 \int d\psi_0^I \int d\tilde\psi_0^I$ with the path integral over $\psi^I$ and
$\tilde \psi^I$ to get the formula 
\begin{equation}\label{finalamp}
	{\mathcal{A}}_g = \int_{{\mathcal{M}}_g}\left\langle\left|\prod_{i=1}^{3g-3}G^-_{int}(\mu_i)\right|^2\right\rangle_{top}\int d^4 x \int d^2\theta \int d^2\tilde{\theta}(P_{\alpha\beta}P^{\alpha\beta})^g.
\end{equation}
where $\langle...\rangle_{top}$ denotes the path integral over all modes of $x^I, x_I, \psi^I, \psi_I$.  Thus, the coupling to the supersymmetric term containing $\int d^4 x R^2 T^{2g-2}$ is precisely the partition function $F_g^B =\int_{{\mathcal{M}}_g}\left\langle\left|\prod_{i=1}^{3g-3}G^-_{int}(\mu_i)\right|^2\right\rangle_{top}$ of the B-model topological string theory at genus $g$ when $g>1$. This same result will now
be shown to also occur when $g=1$.

\subsection{One-loop amplitude}

The non-minimal pure spinor prescription for one-loop amplitudes involving two external states is 
\begin{equation}\label{presone}
{\mathcal{A}_1} =\int_{{\mathcal{M}}_1}\left\langle\left|{\mathcal{N}}b(\mu)\right|^2  ~U~ V(z)\right\rangle,
\end{equation}
where $U$ is the integrated vertex operator for one of the states, 
$$V(z)=\lambda^{\dot{\alpha}}\tilde\lambda^{\dot\beta}A_{\dot\alpha\dot\beta}(x,\theta, \tilde\theta)$$ is the
BRST-invariant unintegrated vertex operator for the other state which is inserted anywhere on the surface, and $A_{\dot\alpha\dot\beta}(x,\theta, \tilde\theta)$ is the bispinor prepotential superfield appearing in (\ref{vcv}). As before, the four-dimensional version of $b^{(a)}$ can be substituted for the $b$ ghost when computing
the one-loop scattering of two anti-self-dual gravitons.

Because of the integration over fermionic zero modes, only the twisted sector will contribute to this one-loop amplitude and the fermionic zero modes of the left-moving variables are
$(\theta^\alpha, \bar\theta^{\dot\alpha}, r_{\dot\alpha})$ and $(p_\alpha, \bar p_{\dot\alpha}, s^{\dot\alpha})$. The two zero modes of $p_\alpha$ must come
from $U$ and $b^{(a)}$ where $b^{(a)}$ of (\ref{bfour}) contributes 
$-{{\bar{\lambda}_{\dot\alpha}}\over {2 \bar\lambda_{\dot\gamma}\lambda^{\dot\gamma}}} \Big( \partial x^\mu \sigma_\mu^{\dot\alpha\beta} p_\beta \Big)$, the $\theta^\alpha$ zero modes come from $V$ and $U$, and the remaining zero modes come from the regulator ${\mathcal{N}}$.

To relate this one-loop amplitude with the one-loop topological partition function, use the OPE
$b(y) J_g(z) \to (y-z)^{-1} b(z)$, where $J_g=w_{\ubbm\alpha} \lambda^{\ubbm\alpha}+ r_{\ubbm\alpha} s^{\ubbm\alpha}$ is the ghost-number current, to write the $b^{(a)}$ ghost as the contour integral of $b^{(a)}$ around $J_g$, i.e.
\begin{equation}
b^{(a)}(\mu) = [\oint b^{(a)}, J_g (\mu)].
\end{equation}
The contour integral can be pulled off of $J_g$ and
since the commutator of $\oint b^{(a)}$ with $U$ and ${\mathcal{N}}$ does not
contain enough zero modes of $p_\alpha$, the only contribution
comes from the commutator with $V$. Performing the same
operation with the right-moving $\tilde b^{(a)}$ ghost, one can write the one-loop amplitude as
\begin{equation}\label{presoneb}
{\mathcal{A}_1} =\int_{{\mathcal{M}}_1}\left\langle\left|{\mathcal{N}} J_g(\mu)\right|^2  ~U~ \oint b^{(a)} \oint \tilde b^{(a)} V(z)\right\rangle.
\end{equation}
In the gauge where $b^{(a)}$ and $\tilde b^{(a)}$ have no double poles with $V =\lambda^{\dot{\alpha}}\tilde\lambda^{\dot\beta}A_{\dot\alpha\dot\beta}$, the superfields $P^{\alpha\beta}$ and
$A_{\dot\alpha\dot\beta}$ are related by
\begin{equation}\label{rel}
P^{\alpha\beta}=(\sigma^\mu)^{\alpha\dot\alpha} (\sigma^\nu)^{\beta\dot\beta}\partial_\mu\partial_\nu A_{\dot\alpha\dot\beta}.
\end{equation} 
So after integrating over the fermionic zero modes of $(\theta^\alpha, \tilde\theta^\alpha)$ and $(d_\alpha, \tilde d_\alpha)$, and using (\ref{rel}), 
(\ref{presoneb}) reduces to
\begin{equation}\label{presonec}
{\mathcal{A}_1} =\int_{{\mathcal{M}}_1}\left\langle\left| {\mathcal{N}} J_g (\mu)\right|^2\right\rangle \int d^4 x \int d^2\theta \int d^2\tilde{\theta}(P_{\alpha\beta}P^{\alpha\beta})
\end{equation}
where $\left\langle~~\right\rangle$ denotes the functional integral over all worldsheet variables except for the four-dimensional variables $(x^\mu, \theta^\alpha, \tilde\theta^\alpha, p_\alpha, \tilde p_\alpha)$.

Finally, one can relate $\int_{{\mathcal{M}}_1}\left\langle\left|{\mathcal{N}} J_g (\mu)\right|^2\right\rangle$ to the one-loop topological partition function by using the
field redefinition of (\ref{change1}) and
shifting $w_{\dot\alpha}$ and $\bar w^{\dot\alpha}$ as in (\ref{shift}). Since $\psi_I$ and $\psi^I$ carry
ghost-number charge $+1$ and $-1$,
\begin{equation}
J_g = K + \psi^I\psi_I
\end{equation}
where $K = J_g - \psi^I \psi_I$ is independent of $(\psi^I, \psi_I)$ and is constructed from the pure spinor variables and $(\chi^I, \chi_I)$. Note that $J_g$ satisfies the OPE  $J_g(y) J_g(z) \to 3 (y-z)^{-2}$ , so
the OPE of $K$ with $K$ has no double pole. One can therefore define $K = \partial \sigma$ where $\sigma$ is a null boson and construct new fermionic variables 
\begin{equation}
\zeta^I = \psi^I e^{+{1\over 3}\sigma}, \quad \zeta_I = \psi_I e^{-{1\over 3}\sigma},
\end{equation}
which satisfy the same OPE's and twistings as $\psi^I$ and $\psi_I$, and satisfy
\begin{equation}
J_g = \zeta^I\zeta_I.
\end{equation}
The integration over all fermionic and bosonic variables in (\ref{presonec}) except for the internal $(x^I, x_I, \zeta^I, \zeta_I, \tilde\zeta^I, \tilde\zeta_I)$ variables cancels out, so the one-loop amplitude can be expressed as 
\begin{equation}\label{presoned}
{\mathcal{A}_1} =F_1^B \int d^4 x \int d^2\theta \int d^2\tilde{\theta}(P_{\alpha\beta}P^{\alpha\beta})
\end{equation}
where $F_1^B$ is the one-loop partition function of the topological B-model \cite{bc}
$$F_1^B = \int_{{\mathcal{M}}_1}\left\langle\left| \zeta^I\zeta_I  (\mu)\right|^2\right\rangle_{top}$$
and $\left\langle ~~~\right\rangle_{top}$ denotes integration over the internal variables $(x^I, x_I, \zeta^I, \zeta_I, \tilde\zeta^I, \tilde\zeta_I)$.

\subsection{Other topological amplitudes}

In the previous subsection, the Type IIB scattering of $2g-2$ anti-self-dual graviphotons and 2 anti-self-dual gravitons was computed using
the non-minimal pure spinor prescription of (\ref{pres}) with $b^{(a)}$ and $\tilde b^{(a)}$ ghosts to obtain the amplitude
\begin{equation}
	F^B_g\int d^4x\ d^2\theta d^2\tilde{\theta}\ (P_{\alpha\beta}P^{\alpha\beta})^g.
\end{equation}
where $F^B_g=\int_{{\mathcal{M}}_g}\left\langle\left|\prod_{i=1}^{3g-3}G^-_{int}(\mu_i)\right|^2\right\rangle_{top}$ is the topological B-model partition function.
Using identical reasoning, one can compute the Type IIB scattering of $2g-2$ self-dual graviphotons and 2 self-dual gravitons using
the prescription of (\ref{pres}) with $b^{(c)}$ and $\tilde b^{(c)}$ ghosts. In this case, one
restricts to patches where $\bar\lambda_\alpha \lambda^\alpha \neq 0$, and the resulting amplitude is
\begin{equation}
	\bar F^B_g\int d^4x\ d^2\bar\theta d^2\tilde{\bar\theta}\ ( P_{\dot\alpha\dot\beta}P^{\dot\alpha\dot\beta})^g
\end{equation}
where $\bar F^B_g$ is the complex conjugate of $F^B_g$ defined by
\begin{equation}
	\bar F^B_g=\int_{{\mathcal{M}}_g}\left\langle\left|\prod_{i=1}^{3g-3}G^+_{int}(\mu_i)\right|^2\right\rangle_{top},
\end{equation}
$G^+_{int} \equiv \psi_I\partial x^I$ is the contribution from the $b^{(c)}$ ghost, and   $\psi_I=(\bar{\lambda}^\beta\lambda_\beta)^{-1}\bar{\lambda}^\alpha p_{\alpha I}$.

One can also compute the Type IIB scattering of $2g-2$ Ramond-Ramond hypermultiplet scalars and 2 NS-NS hypermultiplet scalars with the prescription of (\ref{pres}) if one uses either $b^{(a)}$ ghosts in the left-moving sector and
$\tilde b^{(c)}$ ghosts in the right-moving sector or $b^{(c)}$ ghosts in the left-moving sector and
$\tilde b^{(a)}$ ghosts in the right-moving sector. In the first case, the amplitude is
\begin{equation}\label{qqaction}
	F^A_g\int d^4x\int d^2\theta \int d^2\tilde{\bar{\theta}}\ (Q_{\alpha\dot{\beta}}Q^{\alpha\dot{\beta}})^g.
\end{equation}
where $F^A_g$ is the partition function at genus $g$ for the A model topological string defined as
\begin{equation}
	F^A_g=\int_{{\mathcal{M}}_g}\left\langle\prod_{i=1}^{3g-3}G^-_{int}(\mu_i)\prod_{j=1}^{3g-3}\tilde{G}^+_{int}(\bar{\mu}_j)\right\rangle_{top}.
\end{equation}
In the second case, one obtains the complex conjugate of (\ref{qqaction}) which is
\begin{equation}
	\bar{F}^A_g\int d^4x\int d^2\bar{\theta}\int d^2\tilde{\theta}\ ({Q}_{\dot{\alpha}\beta}{Q}^{\dot{\alpha}\beta})^g,
\end{equation}
where
\begin{equation}
	\bar F^A_g=\int_{{\mathcal{M}}_g}\left\langle\prod_{i=1}^{3g-3}G^+_{int}(\mu_i)\prod_{j=1}^{3g-3}\tilde{G}^-_{int}(\bar{\mu}_j)\right\rangle_{top}.
\end{equation}

For type IIA topological amplitudes, the results are complementary and the only difference comes from the opposite chirality of the right-moving sector as compared with type IIB. For the anti-self-dual graviphoton amplitude, the right-moving $b$-ghost now contributes with $(\tilde{\bar{\lambda}}_{\dot{\alpha}}\tilde{\lambda}^{\dot{\alpha}})^{-1}\bar{\partial}x^I\tilde{p}^{\dot{\alpha}}_I$ and the appropriate change of variables that has to be performed to make contact with the topological string description of the internal model is
\begin{equation}
\tilde{\psi}_I=(\tilde{\bar{\lambda}}_{\dot{\beta}}\tilde{\lambda}^{\dot{\beta}})^{-1}\tilde{\bar{\lambda}}_{\dot{\alpha}}\tilde{p}^{\dot{\alpha}}_I,\hspace{1cm}\tilde{\chi}_I=\tilde{\lambda}_{\dot{\alpha}}\tilde{p}^{\dot{\alpha}}_I,
\end{equation}
\begin{equation}
\tilde{\psi}^I=\tilde{\lambda}_{\dot{\alpha}}\tilde{\theta}^{\dot{\alpha}I},\hspace{1cm}\tilde{\chi}^I=(\tilde{\bar{\lambda}}_{\dot{\beta}}\tilde{\lambda}^{\dot{\beta}})^{-1}\tilde{\bar{\lambda}}_{\dot{\alpha}}\tilde{\theta}^{\dot{\alpha}I},
\end{equation}
The effect of this is to change the relative topological twisting in the right-moving sector of the internal model, and the corresponding terms in the Type IIA effective action are
\begin{equation}
\begin{array}{ccl}
	S &=& \displaystyle F^A_g\int d^4x\int d^2\theta \int d^2\tilde{\theta}\ (P_{\alpha\beta}P^{\alpha\beta})^g+\bar{F}^A_g\int d^4x\int d^2\bar{\theta}\int d^2\tilde{\bar{\theta}}\ (P_{\dot{\alpha}\dot{\beta}}P^{\dot{\alpha}\dot{\beta}})^g\\
	 & & \displaystyle+F^B_g\int d^4x\int d^2\theta \int d^2\tilde{\bar{\theta}}\ (Q_{\alpha\dot{\beta}}Q^{\alpha\dot{\beta}})^g+\bar{F}^B_g\int d^4x\int d^2\bar{\theta}\int d^2\tilde{\theta}\ ({Q}_{\dot{\alpha}\beta}{Q}^{\dot{\alpha}\beta})^g.
\end{array}
\end{equation}

\section{Concluding Remarks}\label{sec:conclusions}

In this paper, certain multiloop amplitudes in Type II superstring theory compactified on an orbifold to four dimensions were computed using the non-minimal pure spinor
formalism and related to higher genus partition functions of topological string theory. As in the topological amplitude computations using the hybrid formalism, these computations
using the pure spinor formalism preserve manifest $N=2$ $D=4$ supersymmetry and do not require summing over spin structures. Although the pure spinor vertex operators
depend on all 16 $\theta^{\ubbm\alpha}$ and 16 $\tilde\theta^{\ubbm\alpha}$ worldsheet variables, the restriction to patches where $\bar\lambda^{\dot\alpha}\neq 0$ and the construction
of a four-dimensional version of the $b$ ghost simplify the computations. Moreover,  the multiloop computations using the non-minimal pure spinor formalism do not suffer from the subtleties in the hybrid formalism
computations coming from negative-energy chiral bosons.

In future work, it would be very interesting to use the simplified four-dimensional version of the $b$ ghost to compute other non-topological amplitudes with the pure spinor formalism
compactified on an orbifold. It would also be interesting to generalize the construction of the four-dimensional $b$ ghost and the computation of scattering amplitudes using
the pure spinor formalism to compactifications on general Calabi-Yau backgrounds that
preserve $N=2$ $D=4$ supersymmetry.

\vskip 15pt
{\bf Acknowledgments:}
We would like to thank Kumar Narain and Edward Witten for useful discussions and ICTP Trieste for their hospitality.
LAY acknowledges CNPq grant 141708/2016-6 for financial support and NB acknowledges FAPESP grants 2016/01343-7 and 2014/18634-9 and CNPq grant 300256/94-9 for partial financial support.

\appendix

\section{BRST Equivalence of Four-Dimensional $b$ Ghost}\label{app:fourb}

In this section of the appendix, the $b$ ghost of  (\ref{bghostnm}) will be shown to satisfy $b=b^{(c)} + Q\Lambda^{(c)}$  for some $\Lambda^{(c)}$. Note that terms in $b$ have denominators which are powers of $\llbar{\ubbm\alpha}$, while those in $b^{(c)}$ have powers of $\llbar{\alpha}$. The strategy will be to manipulate terms in $b$ order by order in $(\llbar{\ubbm\alpha})^{-1}$ in such a way as to trade them for terms with $(\llbar{\alpha})^{-1}$. This will produce some BRST trivial terms and extra non-trivial terms which will be canceled by expressions coming from manipulations at next order in the analysis. In the end, all non-trivial terms will cancel each other due to the relations (\ref{descent1}) and (\ref{descent2}). 

The convention used for spinor indices is the following: $\alpha$ denotes chiral spinors in four dimensions, and $\alpha'$ denotes any of the other components in the ten-dimensional quantity, $\alpha'=(\dot{\alpha},\alpha I, \dot{\alpha}I)$ where the position of $I$ depends on the chirality of the ten-dimensional spinor.

\subsubsection*{First term}

To analyze the first term in the $b$ ghost of (\ref{bghostnm}), use the  relation
\begin{equation}
Q(\bar{\lambda}_\alpha\bar{\lambda}_{\alpha'}H^{[\alpha,\alpha']})=\llbar{\alpha}\bar{\lambda}_{\alpha'}G^{\alpha'}-\bar{\lambda}_{\alpha'}\lambda^{\alpha'}\bar{\lambda}_\alpha G^\alpha-\bar{\lambda}_{\alpha'}r_\alpha H^{[\alpha,\alpha']}-\bar{\lambda}_\alpha r_{\alpha'}H^{[\alpha,\alpha']}
\end{equation}
to express
\begin{equation}\label{lg}
\bar{\lambda}_{\alpha'}G^{\alpha'}={1\over\llbar{\alpha}}\Big[Q(\bar{\lambda}_\alpha\bar{\lambda}_{\alpha'}H^{[\alpha,\alpha']})+\llbar{\alpha'}\bar{\lambda}_\alpha G^\alpha+\bar{\lambda}_{\alpha'}r_\alpha H^{[\alpha,\alpha']}+\bar{\lambda}_\alpha r_{\alpha'}H^{[\alpha,\alpha']}\Big],
\end{equation}
where the patch in pure spinor space is $\llbar{\alpha}\neq 0$.
Then the first term in the $b$-ghost can be written as
\begin{equation*}
b_1={\bar{\lambda}_{\ubbm\alpha}G^{\ubbm\alpha}\over\llbar{\ubbm\alpha}}={1\over\llbar{\ubbm\alpha}}(\bar{\lambda}_\alpha G^\alpha+\bar{\lambda}_{\alpha'}G^{\alpha'})=
\end{equation*}
\begin{equation*}
={\bar{\lambda}_\alpha G^\alpha\over\llbar{\alpha}}+Q\left({\bar{\lambda}_\alpha\bar{\lambda}_{\alpha'}H^{[\alpha,\alpha']}\over(\llbar{\alpha})(\llbar{\ubbm\alpha})}\right)-\bar{\lambda}_\alpha\bar{\lambda}_{\alpha'}H^{[\alpha,\alpha']}Q\left({1\over(\llbar{\alpha})(\llbar{\ubbm\alpha})}\right)
\end{equation*}
\begin{equation}\label{b1}
+{\bar{\lambda}_\alpha r_{\alpha'}H^{[\alpha,\alpha']}-\bar{\lambda}_{\alpha'}r_\alpha H^{[\alpha',\alpha]}\over(\llbar{\alpha})(\llbar{\ubbm\alpha})}.
\end{equation}

\subsubsection*{Second term}
Similarly, from the relation
\begin{equation*}
Q(\bar{\lambda}_\alpha\bar{\lambda}_{\beta'}r_\gamma K^{[\alpha,\beta',\gamma]}) =-\llbar{\alpha}\bar{\lambda}_{\beta'}r_\gamma H^{[\beta',\gamma]}-\llbar{\beta'}\bar{\lambda}_\alpha r_\gamma H^{[\gamma,\alpha]}-r_\gamma\lambda^\gamma\bar{\lambda}_\alpha\bar{\lambda}_{\beta'}H^{[\alpha,\beta']}
\end{equation*}
\begin{equation}
-\bar{\lambda}_{\beta'}r_\alpha r_\gamma K^{[\alpha,\beta',\gamma]}-\bar{\lambda}_\alpha r_{\beta'}r_\gamma K^{[\alpha,\beta',\gamma]}
\end{equation}
it follows that
\begin{equation*}
\bar{\lambda}_{\beta'}r_\gamma H^{[\beta',\gamma]}={1\over\llbar{\alpha}}\left[\llbar{\beta'}\bar{\lambda}_\alpha r_\gamma H^{[\alpha,\gamma]}-r_\gamma\lambda^\gamma\bar{\lambda}_\alpha\bar{\lambda}_{\beta'}H^{[\alpha,\beta']}\right.
\end{equation*}
\begin{equation}\label{lrh1}
\left.-\bar{\lambda}_{\beta'}r_\alpha r_\gamma K^{[\alpha,\beta',\gamma]}-\bar{\lambda}_\alpha r_{\beta'}r_\gamma K^{[\alpha,\beta',\gamma]}-Q(\bar{\lambda}_\alpha\bar{\lambda}_{\beta'}r_\gamma K^{[\alpha,\beta',\gamma]})\right].
\end{equation}

Again, from
\begin{equation*}
Q(\bar{\lambda}_\alpha\bar{\lambda}_{\beta'}r_{\gamma'} K^{[\alpha,\beta',\gamma']}) =-\llbar{\alpha}\bar{\lambda}_{\beta'}r_{\gamma'} H^{[\beta',\gamma']}-\llbar{\beta'}\bar{\lambda}_\alpha r_{\gamma'} H^{[\gamma',\alpha]}-r_{\gamma'}\lambda^{\gamma'}\bar{\lambda}_\alpha\bar{\lambda}_{\beta'}H^{[\alpha,\beta']}
\end{equation*}
\begin{equation}
-\bar{\lambda}_{\beta'}r_\alpha r_{\gamma'} K^{[\alpha,\beta',\gamma']}-\bar{\lambda}_\alpha r_{\beta'}r_{\gamma'} K^{[\alpha,\beta',\gamma']},
\end{equation}
one can express
\begin{equation*}
\bar{\lambda}_{\beta'}r_{\gamma'} H^{[\beta',\gamma']}={1\over\llbar{\alpha}}\left[\llbar{\beta'}\bar{\lambda}_\alpha r_{\gamma'} H^{[\alpha,\gamma']}-r_{\gamma'}\lambda^{\gamma'}\bar{\lambda}_\alpha\bar{\lambda}_{\beta'}H^{[\alpha,\beta']}\right.
\end{equation*}
\begin{equation}\label{lrh2}
\left.-\bar{\lambda}_{\beta'}r_\alpha r_{\gamma'} K^{[\alpha,\beta',\gamma']}-\bar{\lambda}_\alpha r_{\beta'}r_{\gamma'} K^{[\alpha,\beta',\gamma']}-Q(\bar{\lambda}_\alpha\bar{\lambda}_{\beta'}r_{\gamma'} K^{[\alpha,\beta',\gamma']})\right].
\end{equation}

Expanding
\begin{equation}
-\bar{\lambda}_{\ubbm\alpha}r_{\ubbm\beta}H^{[\ubbm\alpha,\ubbm\beta]}=-\bar{\lambda}_\alpha r_\beta H^{[\alpha,\beta]}-\bar{\lambda}_\alpha r_{\beta'}H^{[\alpha,\beta']}-\bar{\lambda}_{\alpha'}r_\beta H^{[\alpha',\beta]}-\bar{\lambda}_{\alpha'}r_{\beta'}H^{[\alpha',\beta']}
\end{equation}

and denoting
\begin{equation}
A=-\bar{\lambda}_\alpha r_\beta H^{[\alpha,\beta]}-\bar{\lambda}_{\alpha'}r_\beta H^{[\alpha',\beta]},
\end{equation}
\begin{equation}
B=-\bar{\lambda}_\alpha r_{\beta'}H^{[\alpha,\beta']}-\bar{\lambda}_{\alpha'}r_{\beta'}H^{[\alpha',\beta']},
\end{equation}

the second term in the $b$-ghost can be written as
\begin{equation}
b_2={A+B\over(\llbar{\ubbm\alpha})^2}.
\end{equation}
Plugging in equations (\ref{lrh1}) and (\ref{lrh2}) in $A$ and $B$, respectively, one gets
\begin{equation*}
A={1\over\llbar{\alpha}}\left[-\llbar{\ubbm\alpha}\bar{\lambda}_\alpha r_\beta H^{[\alpha,\beta]}+r_\gamma\lambda^\gamma\bar{\lambda}_\alpha\bar{\lambda}_{\beta'}H^{[\alpha,\beta']}\right.
\end{equation*}
\begin{equation}
\left.+\bar{\lambda}_{\beta'}r_\alpha r_\gamma K^{[\alpha,\beta',\gamma]}+\bar{\lambda}_\alpha r_{\beta'}r_\gamma K^{[\alpha,\beta',\gamma]}+Q\Big(\bar{\lambda}_\alpha\bar{\lambda}_{\beta'}r_\gamma K^{[\alpha,\beta',\gamma]}\Big)\right],
\end{equation}
\begin{equation*}
B={1\over\llbar{\alpha}}\left[-\llbar{\ubbm\alpha}\bar{\lambda}_\alpha r_{\beta'}H^{[\alpha,\beta']}+r_{\gamma'}\lambda^{\gamma'}\bar{\lambda}_\alpha\bar{\lambda}_{\beta'}H^{[\alpha,\beta']}\right.
\end{equation*}
\begin{equation}
\left.+\bar{\lambda}_{\beta'}r_\alpha r_{\gamma'}K^{[\alpha,\beta',\gamma']}+\bar{\lambda}_\alpha r_{\beta'}r_{\gamma'}K^{[\alpha,\beta',\gamma']}+Q\Big(\bar{\lambda}_\alpha\bar{\lambda}_{\beta'}r_{\gamma'}K^{[\alpha,\beta',\gamma']}\Big)\right].
\end{equation}\\

The first term in $B$ after being multiplied by $(\llbar{\ubbm\alpha})^{-2}$ cancels one of the last terms in (\ref{b1}). The first term in $A$ times $(\llbar{\ubbm\alpha})^{-2}$ combines with the last term in (\ref{b1}) to produce $A(\llbar{\alpha})^{-1}(\llbar{\ubbm\alpha})^{-1}$, and this is
\begin{equation*}
-{\bar{\lambda}_\alpha r_\beta H^{[\alpha,\beta]}\over(\llbar{\alpha})^2}+{1\over(\llbar{\alpha})^2\llbar{\ubbm\alpha}}\left[r_\gamma\lambda^\gamma\bar{\lambda}_\alpha\bar{\lambda}_{\beta'}H^{[\alpha,\beta']}\right.
\end{equation*}
\begin{equation}
\left.+\bar{\lambda}_{\beta'}r_\alpha r_\gamma K^{[\alpha,\beta',\gamma]}+\bar{\lambda}_\alpha r_{\beta'}r_\gamma K^{[\alpha,\beta',\gamma]}+Q\Big(\bar{\lambda}_\alpha\bar{\lambda}_{\beta'}r_\gamma K^{[\alpha,\beta',\gamma]}\Big)\right].
\end{equation}

The second term in $A$ combines with the second term in $B$ to produce
\begin{equation}
{r_{\ubbm\alpha}\lambda^{\ubbm\alpha}\bar{\lambda}_\alpha\bar{\lambda}_{\beta'}H^{[\alpha,\beta']}\over\llbar{\alpha}(\llbar{\ubbm\alpha})^2}.
\end{equation}
This term and the one in $A(\llbar{\alpha})^{-1}(\llbar{\ubbm\alpha})^{-1}$, that is,
\begin{equation}
{r_\gamma\lambda^\gamma\bar{\lambda}_\alpha\bar{\lambda}_{\beta'}H^{[\alpha,\beta']}\over(\llbar{\alpha})^2\llbar{\ubbm\alpha}},
\end{equation}
will cancel the entire term containing $Q\left({1\over(\llbar{\alpha})(\llbar{\ubbm\alpha})}\right)$, as can be seen from
\begin{equation}
Q\left({1\over(\llbar{\alpha})(\llbar{\ubbm\alpha})}\right)={r_\alpha\lambda^\alpha\over(\llbar{\alpha})^2\llbar{\ubbm\alpha}}+{r_{\ubbm\alpha}\lambda^{\ubbm\alpha}\over\llbar{\alpha}(\llbar{\ubbm\alpha})^2}.
\end{equation}\\

Summarizing all this,
\begin{equation*}
b_1+b_2={\bar{\lambda}_\alpha G^\alpha\over\llbar{\alpha}}-{\bar{\lambda}_\alpha r_\beta H^{[\alpha,\beta]}\over(\llbar{\alpha})^2}
\end{equation*}
\begin{equation*}
+Q\bigg({\bar{\lambda}_\alpha\bar{\lambda}_{\beta'}H^{[\alpha,\beta']}\over(\llbar{\alpha})(\llbar{\ubbm\alpha})}+{\bar{\lambda}_\alpha\bar{\lambda}_{\beta'}r_{\ubbm\gamma}K^{[\alpha,\beta',\ubbm\gamma]}\over\llbar{\alpha}(\llbar{\ubbm\alpha})^2}+{\bar{\lambda}_\alpha\bar{\lambda}_{\beta'}r_\gamma K^{[\alpha,\beta',\gamma]}\over(\llbar{\alpha})^2\llbar{\ubbm\alpha}}\bigg)
\end{equation*}
\begin{equation*}
+{\bar{\lambda}_{\beta'}r_\alpha r_{\ubbm\gamma}K^{[\alpha,\beta',\ubbm\gamma]}+\bar{\lambda}_\alpha r_{\beta'}r_{\ubbm\gamma}K^{[\alpha,\beta'.\ubbm\gamma]}\over\llbar{\alpha}(\llbar{\ubbm\alpha})^2}+{\bar{\lambda}_{\beta'}r_\alpha r_{\gamma}K^{[\alpha,\beta',\gamma]}+\bar{\lambda}_\alpha r_{\beta'}r_{\gamma}K^{[\alpha,\beta',\gamma]}\over(\llbar{\alpha})^2\llbar{\ubbm\alpha}}
\end{equation*}
\begin{equation}\label{b12}
-\bar{\lambda}_\alpha\bar{\lambda}_{\beta'}r_{\ubbm\gamma}K^{[\alpha,\beta',\ubbm\gamma]}Q\bigg({1\over\llbar{\alpha}(\llbar{\ubbm\alpha})^2}\bigg)-\bar{\lambda}_\alpha\bar{\lambda}_{\beta'}r_{\gamma}K^{[\alpha,\beta',\gamma]}Q\bigg({1\over(\llbar{\alpha})^2\llbar{\ubbm\alpha}}\bigg).
\end{equation}

\subsubsection*{Third term}
From the expression
\begin{equation*}
Q(\bar{\lambda}_\alpha\bar{\lambda}_{\beta'}r_{\hat{\gamma}}r_{\hat{\delta}}L^{[\alpha,\beta',\hat{\gamma},\hat{\delta}]})=\llbar{\alpha}\bar{\lambda}_{\beta'}r_{\hat{\gamma}}r_{\hat{\delta}}K^{[\beta',\hat{\gamma},\hat{\delta}]}-\llbar{\beta'}\bar{\lambda}_\alpha r_{\hat{\gamma}}r_{\hat{\delta}}K^{[\alpha,\hat{\gamma},\hat{\delta}]}
\end{equation*}
\begin{equation*}
+r_{\hat{\gamma}}\lambda^{\hat{\gamma}}\bar{\lambda}_\alpha\bar{\lambda}_{\beta'}r_{\hat{\delta}}K^{[\alpha,\beta',\hat{\delta}]}+r_{\hat{\delta}}\lambda^{\hat{\delta}}\bar{\lambda}_\alpha\bar{\lambda}_{\beta'}r_{\hat{\gamma}}K^{[\alpha,\beta',\hat{\gamma}]}
\end{equation*}
\begin{equation}
-\bar{\lambda}_{\beta'}r_\alpha r_{\hat{\gamma}}r_{\hat{\delta}}L^{[\alpha,\beta',\hat{\gamma},\hat{\delta}]} -\bar{\lambda}_{\alpha}r_{\beta'} r_{\hat{\gamma}}r_{\hat{\delta}}L^{[\alpha,\beta',\hat{\gamma},\hat{\delta}]},
\end{equation}
where the equation holds for every formula obtained by replacing indices like $\hat{\gamma}$ by $\gamma$ or $\gamma'$, one can write
\begin{equation*}
\bar{\lambda}_{\beta'}r_{\hat{\gamma}}r_{\hat{\delta}}K^{[\beta',\hat{\gamma},\hat{\delta}]} = {1\over\llbar{\alpha}}\left[\llbar{\beta'}\bar{\lambda}_\alpha r_{\hat{\gamma}}r_{\hat{\delta}}K^{[\alpha,\hat{\gamma},\hat{\delta}]}-r_{\hat{\gamma}}\lambda^{\hat{\gamma}}\bar{\lambda}_\alpha\bar{\lambda}_{\beta'}r_{\hat{\delta}}K^{[\alpha,\beta',\hat{\delta}]}-r_{\hat{\delta}}\lambda^{\hat{\delta}}\bar{\lambda}_\alpha\bar{\lambda}_{\beta'}r_{\hat{\gamma}}K^{[\alpha,\beta',\hat{\gamma}]}\right.
\end{equation*}
\begin{equation}\label{lrrk1}
\left.+\bar{\lambda}_{\beta'}r_\alpha r_{\hat{\gamma}}r_{\hat{\delta}}L^{[\alpha,\beta',\hat{\gamma},\hat{\delta}]} +\bar{\lambda}_{\alpha}r_{\beta'} r_{\hat{\gamma}}r_{\hat{\delta}}L^{[\alpha,\beta',\hat{\gamma},\hat{\delta}]}+Q(\bar{\lambda}_\alpha\bar{\lambda}_{\beta'}r_{\hat{\gamma}}r_{\hat{\delta}}L^{[\alpha,\beta',\hat{\gamma},\hat{\delta}]})\right].
\end{equation}
For the particular case where $(\hat{\gamma},\hat{\delta})=(\gamma,\delta)$ the previous formula reduces to
\begin{equation}\label{lrrk2}
\bar{\lambda}_{\beta'}r_\gamma r_\delta K^{[\beta',\gamma,\delta]}=-{2\over\llbar{\alpha}}r_\gamma\lambda^\gamma\bar{\lambda}_\alpha\bar{\lambda}_{\beta'}r_\delta K^{[\alpha,\beta',\delta]}.
\end{equation}

The third term in the $b$-ghost is
\begin{equation}
b_3=-{\bar{\lambda}_{\ubbm\alpha}r_{\ubbm\beta}r_{\ubbm\gamma}K^{[\ubbm\alpha,\ubbm\beta,\ubbm\gamma]}\over(\llbar{\ubbm\alpha})^3}
\end{equation}
which can be written as $b_3=(C+D+E)(\llbar{\ubbm\alpha})^{-3}$,
where
\begin{equation}
C=-\bar{\lambda}_\alpha r_{\ubbm\beta}r_{\gamma'}K^{[\alpha,\ubbm\beta,\gamma']}-\bar{\lambda}_{\alpha'}r_{\ubbm\beta}r_{\gamma'}K^{[\alpha',\ubbm\beta,\gamma']},
\end{equation}
\begin{equation}
D=-\bar{\lambda}_\alpha r_{\beta'}r_\gamma K^{[\alpha,\beta',\gamma]}-\bar{\lambda}_{\alpha'}r_{\beta'}r_{\gamma}K^{[\alpha',\beta',\gamma]},
\end{equation}
\begin{equation}
E=-\bar{\lambda}_{\alpha'}r_\beta r_\gamma K^{[\alpha',\beta,\gamma]}.
\end{equation}

Focusing first on terms containing expressions of the form $\bar{\lambda}_{\hat{\alpha}}r_{\hat{\beta}}r_{\hat{\gamma}}K^{[\hat{\alpha},\hat{\beta},\hat{\gamma}]}$ after making the substitutions (\ref{lrrk1}) and (\ref{lrrk2}), one gets several such terms.
The terms in $(C+D)(\llbar{\ubbm\alpha})^{-3}$ of this form are
\begin{equation}
-{\bar{\lambda}_\alpha r_{\ubbm\beta}r_{\gamma'}K^{[\alpha,\ubbm\beta,\gamma']}\over\llbar{\alpha}(\llbar{\ubbm\alpha})^2} -{\bar{\lambda}_\alpha r_{\beta'}r_{\gamma}K^{[\alpha,\beta',\gamma]}\over\llbar{\alpha}(\llbar{\ubbm\alpha})^2}
\end{equation}
which, together with terms having same denominator in (\ref{b12}), give
\begin{equation}
{D\over\llbar{\alpha}(\llbar{\ubbm\alpha})^2}-{\bar{\lambda}_{\beta'}r_\alpha r_\gamma K^{[\beta',\alpha,\gamma]}\over\llbar{\alpha}(\llbar{\ubbm\alpha})^2}.
\end{equation}
The first term in the last formula contains
\begin{equation}
-{\bar{\lambda}_\alpha r_{\beta'}r_\gamma K^{[\alpha,\beta',\gamma]}\over(\llbar{\alpha})^2\llbar{\ubbm\alpha}}
\end{equation}
which kills one term in (\ref{b12}) with the same denominator. At the end one has in $b_1+b_2+b_3$,
\begin{equation}
-\bar{\lambda}_{\alpha'}r_\beta r_\gamma K^{[\alpha',\beta,\gamma]}\left({1\over(\llbar{\ubbm\alpha})^3}+{1\over\llbar{\alpha}(\llbar{\ubbm\alpha})^2}+{1\over(\llbar{\alpha})^2\llbar{\ubbm\alpha}}\right).
\end{equation}
Using relation (\ref{lrrk2}) one eliminates all appearance of terms containing $\bar{\lambda}_{\hat{\alpha}}r_{\hat{\beta}}r_{\hat{\gamma}}K^{[\hat{\alpha},\hat{\beta},\hat{\gamma}]}$.

It is not difficult to check that all terms containing expressions like $(r\lambda)\bar{\lambda}\bar{\lambda}rK$ cancel and the remaining terms contain only the operator $L^{[\ubbm\alpha,\ubbm\beta,\ubbm\gamma,\ubbm\delta]}$.
One gets
\begin{equation*}
b_1+b_2+b_3={\bar{\lambda}_\alpha G^\alpha\over\llbar{\alpha}}-{\bar{\lambda}_\alpha r_\beta H^{[\alpha,\beta]}\over(\llbar{\alpha})^2}
\end{equation*}
\begin{equation*}
+Q\left({\bar{\lambda}_\alpha\bar{\lambda}_{\beta'}H^{[\alpha,\beta']}\over(\llbar{\alpha})(\llbar{\ubbm\alpha})}+{\bar{\lambda}_\alpha\bar{\lambda}_{\beta'}r_{\ubbm\gamma}K^{[\alpha,\beta',\ubbm\gamma]}\over\llbar{\alpha}(\llbar{\ubbm\alpha})^2}+{\bar{\lambda}_\alpha\bar{\lambda}_{\beta'}r_\gamma K^{[\alpha,\beta',\gamma]}\over(\llbar{\alpha})^2\llbar{\ubbm\alpha}}\right.
\end{equation*}
\begin{equation*}
\left.-{\bar{\lambda}_\alpha\bar{\lambda}_{\beta'}r_{\ubbm\gamma}r_{\delta'}L^{[\alpha,\beta',\ubbm\gamma,\delta']}+\bar{\lambda}_\alpha\bar{\lambda}_{\beta'}r_{\gamma'}r_\delta L^{[\alpha,\beta',\gamma',\delta]}\over\llbar{\alpha}(\llbar{\ubbm\alpha})^3}-{\bar{\lambda}_\alpha\bar{\lambda}_{\beta'}r_{\gamma'}r_\delta L^{[\alpha,\beta',\gamma',\delta]}\over(\llbar{\alpha})^2(\llbar{\ubbm\alpha})^2}\right)
\end{equation*}\

\begin{equation*}
-{(\bar{\lambda}_\alpha r_{\beta'}+\bar{\lambda}_{\beta'}r_\alpha) r_{\gamma'}r_{\delta'}L^{[\alpha,\beta',\gamma',\delta']}+2(\bar{\lambda}_\alpha r_{\beta'}+\bar{\lambda}_{\beta'}r_\alpha)r_{\gamma'}r_\delta L^{[\alpha,\beta',\gamma',\delta]}\over\llbar{\alpha}(\llbar{\ubbm\alpha})^3}
\end{equation*}
\begin{equation*}
-{(\bar{\lambda}_\alpha r_{\beta'}+\bar{\lambda}_{\beta'}r_\alpha)r_{\gamma'}r_\delta L^{[\alpha,\beta',\gamma',\delta]}\over(\llbar{\alpha})^2(\llbar{\ubbm\alpha})^2}
\end{equation*}
\begin{equation*}
+\left(\bar{\lambda}_\alpha\bar{\lambda}_{\beta'}r_{\ubbm\gamma}r_{\delta'}L^{[\alpha,\beta'\ubbm\gamma,\delta']}+\bar{\lambda}_\alpha\bar{\lambda}_{\beta'}r_{\gamma'}r_\delta L^{[\alpha,\beta',\gamma',\delta]}\right)Q\left({1\over\llbar{\alpha}(\llbar{\ubbm\alpha})^3}\right)
\end{equation*}
\begin{equation}
+\bar{\lambda}_\alpha\bar{\lambda}_{\beta'}r_{\gamma'}r_\delta L^{[\alpha,\beta',\gamma',\delta]}Q\left({1\over(\llbar{\alpha})^2(\llbar{\ubbm\alpha})^2}\right).
\end{equation}

\subsubsection*{Fourth term}
The next term of the $b$-ghost is
\begin{equation}
b_4={\bar{\lambda}_{\ubbm\alpha}r_{\ubbm\beta}r_{\ubbm\gamma}r_{\ubbm\delta}L^{[\ubbm\alpha,\ubbm\beta,\ubbm\gamma,\ubbm\delta]}\over(\llbar{\ubbm\alpha})^4}
\end{equation}
where the operator $L^{[\ubbm\alpha,\ubbm\beta,\ubbm\gamma,\ubbm\delta]}$ obeys $\lambda^{[\ubbm\alpha}L^{\ubbm\beta,\ubbm\gamma,\ubbm\delta,\ubbm\rho]}=0$.
These relations can be written as
\begin{equation}
\lambda^{[\alpha'}L^{\beta',\gamma',\delta',\rho']}=0,
\end{equation}
\begin{equation}
\lambda^{\alpha}L^{[\beta',\gamma',\delta',\rho']}-\lambda^{\beta'}L^{[\alpha,\gamma',\delta',\rho']}+\lambda^{\gamma'}L^{[\alpha,\beta',\delta',\rho']}-\lambda^{\delta'}L^{[\alpha,\beta',\gamma',\rho']}+\lambda^{\rho'}L^{[\alpha,\beta',\gamma',\delta']}=0,
\end{equation}
\begin{equation}
\lambda^{\alpha}L^{[\beta,\gamma',\delta',\rho']}-\lambda^{\beta}L^{[\alpha,\gamma',\delta',\rho']}+\lambda^{\gamma'}L^{[\alpha,\beta,\delta',\rho']}-\lambda^{\delta'}L^{[\alpha,\beta,\gamma',\rho']}+\lambda^{\rho'}L^{[\alpha,\beta,\gamma',\delta']}=0,
\end{equation}
\begin{equation}
\lambda^{\alpha}L^{[\beta,\gamma,\delta',\rho']}-\lambda^\beta L^{[\alpha,\gamma,\delta',\rho']}+\lambda^\gamma L^{[\alpha,\beta,\delta',\rho']}=0,
\end{equation}
which implies
\begin{equation}\label{lrrrl1}
\bar{\lambda}_{\beta'}r_{\gamma'}r_{\delta'}r_{\rho'}L^{[\beta',\gamma',\delta',\rho']}={1\over\bar{\lambda}_\alpha\lambda^\alpha}\Big(\bar{\lambda}_{\beta'}\lambda^{\beta'}\bar{\lambda}_\alpha r_{\gamma'}r_{\delta'}r_{\rho'}L^{[\alpha,\gamma',\delta',\rho']}-3r_{\gamma'}\lambda^{\gamma'}\bar{\lambda}_\alpha\bar{\lambda}_{\beta'}r_{\delta'}r_{\rho'}L^{[\alpha,\beta',\delta',\rho']}\Big),
\end{equation}
\begin{equation*}
\bar{\lambda}_{\gamma'}r_{\beta}r_{\delta'}r_{\rho'}L^{[\beta,\gamma',\delta',\rho']}={1\over\llbar{\alpha}}\Big(-\llbar{\gamma'}\bar{\lambda}_\alpha r_\beta r_{\delta'}r_{\rho'}L^{[\alpha,\beta,\delta',\rho']}
\end{equation*}
\begin{equation}\label{lrrrl2}
+r_\beta\lambda^\beta\bar{\lambda}_\alpha\bar{\lambda}_{\gamma'}r_{\delta'}r_{\rho'}L^{[\alpha,\gamma',\delta',\rho']}-2r_{\rho'}\lambda^{\rho'}\bar{\lambda}_\alpha\bar{\lambda}_{\gamma'}r_\beta r_{\delta'}L^{[\alpha,\beta,\gamma',\delta']}\Big),
\end{equation}
\begin{equation}\label{lrrrl3}
\bar{\lambda}_{\delta'}r_\beta r_\gamma r_{\rho'}L^{[\delta',\beta,\gamma,\rho']}={2\over\llbar{\alpha}}r_\beta\lambda^\beta\bar{\lambda}_\alpha\bar{\lambda}_{\delta'}r_\gamma r_{\rho'}L^{[\alpha,\gamma,\delta',\rho']}.
\end{equation}

The fourth term of the $b$-ghost can be written also keeping track of the four dimensional chiral spinor index $\alpha$, as follows:
\begin{equation}
b_4={X+Y+Z\over(\llbar{\ubbm\alpha})^4}
\end{equation}
where
\begin{equation}
X=\bar{\lambda}_{\alpha'}r_{\beta'}r_{\gamma'}r_{\delta'}L^{[\alpha',\beta',\gamma',\delta']}+\bar{\lambda}_\alpha r_{\beta'}r_{\gamma'}r_{\delta'}L^{[\alpha,\beta',\gamma',\delta']},
\end{equation}
\begin{equation}
Y=3\bar{\lambda}_{\alpha'}r_{\beta}r_{\gamma'}r_{\delta'}L^{[\alpha',\beta,\gamma',\delta']}+3\bar{\lambda}_\alpha r_{\beta}r_{\gamma'}r_{\delta'}L^{[\alpha,\beta,\gamma',\delta']},
\end{equation}
\begin{equation}
Z=3\bar{\lambda}_{\alpha'}r_{\beta}r_{\gamma}r_{\delta'}L^{[\alpha',\beta,\gamma,\delta']}.
\end{equation}

Using relations (\ref{lrrrl1}), (\ref{lrrrl2}) and (\ref{lrrrl3}), one can simplify $b_1+b_2+b_3+b_4$. Let's focus again on terms containing $\bar{\lambda}_{\hat{\alpha}}r_{\hat{\beta}}r_{\hat{\gamma}}r_{\hat{\delta}}L^{[\hat{\alpha},\hat{\beta},\hat{\gamma},\hat{\delta}]}$. From $X$, one gets
\begin{equation}
{\bar{\lambda}_{\alpha}r_{\beta'}r_{\gamma'}r_{\delta'}L^{[\alpha,\beta',\gamma',\delta']}\over\llbar{\alpha}(\llbar{\ubbm\alpha})^3} 
\end{equation}
which just cancels one of the terms in $b_1+b_2+b_3$. From $Y$, one gets
\begin{equation}
{3\bar{\lambda}_\alpha r_\beta r_{\gamma'}r_{\delta'}L^{[\alpha,\beta,\gamma',\delta']}\over\llbar{\alpha}(\llbar{\ubbm\alpha})^3}.
\end{equation}
This adds to two terms in $b_1+b_2+b_3$ to produce
\begin{equation}
{Y\over 3\llbar{\alpha}(\llbar{\ubbm\alpha})^3}.
\end{equation}
Using (\ref{lrrrl2}) it is seen that this expression contains
\begin{equation}
{\bar{\lambda}_\alpha r_\beta r_{\gamma'}r_{\delta'}L^{[\alpha,\beta,\gamma',\delta']}\over(\llbar{\alpha})^2(\llbar{\ubbm\alpha})^2}
\end{equation}
which cancels another term in $b_1+b_2+b_3$. What is left is just
\begin{equation}
\bar{\lambda}_{\alpha'}r_\beta r_\gamma r_{\delta'}L^{[\alpha',\beta,\gamma,\delta']}\bigg({3\over(\llbar{\ubbm\alpha})^4}+{2\over(\llbar{\alpha})(\llbar{\ubbm\alpha})^3}+{1\over(\llbar{\alpha})^2(\llbar{\ubbm\alpha})^2}\bigg).
\end{equation}
Using equation (\ref{lrrrl3}) one eliminates all appearances of terms containing $\bar{\lambda}_{\hat{\alpha}}r_{\hat{\beta}}r_{\hat{\gamma}}r_{\hat{\delta}}L^{[\hat{\alpha},\hat{\beta},\hat{\gamma},\hat{\delta}]}$. One ends up with terms having in the numerator expressions like $(r\lambda)\bar{\lambda}\bar{\lambda}rrL$. Collecting all these terms, it is easy to see that they all cancel out.

\subsubsection*{Result}
So the final result for $b=b_1+b_2+b_3+b_4 + s^{\ubbm\alpha}\partial\bar{\lambda}_{\ubbm\alpha}$ is
\begin{equation}
b={\bar{\lambda}_\alpha G^\alpha\over\llbar{\alpha}}-{\bar{\lambda}_\alpha r_\beta H^{[\alpha,\beta]}\over(\llbar{\alpha})^2} +s^{\ubbm\alpha}\partial\bar{\lambda}_{\ubbm\alpha}
\end{equation}
\begin{equation*}
+Q\left({\bar{\lambda}_\alpha\bar{\lambda}_{\beta'}H^{[\alpha,\beta']}\over(\llbar{\alpha})(\llbar{\ubbm\alpha})}+{\bar{\lambda}_\alpha\bar{\lambda}_{\beta'}r_{\ubbm\gamma}K^{[\alpha,\beta',\ubbm\gamma]}\over\llbar{\alpha}(\llbar{\ubbm\alpha})^2}+{\bar{\lambda}_\alpha\bar{\lambda}_{\beta'}r_\gamma K^{[\alpha,\beta',\gamma]}\over(\llbar{\alpha})^2\llbar{\ubbm\alpha}}\right.
\end{equation*}
\begin{equation*}
\left.-{\bar{\lambda}_\alpha\bar{\lambda}_{\beta'}r_{\ubbm\gamma}r_{\delta'}L^{[\alpha,\beta',\ubbm\gamma,\delta']}+\bar{\lambda}_\alpha\bar{\lambda}_{\beta'}r_{\gamma'}r_\delta L^{[\alpha,\beta',\gamma',\delta]}\over\llbar{\alpha}(\llbar{\ubbm\alpha})^3}-{\bar{\lambda}_\alpha\bar{\lambda}_{\beta'}r_{\gamma'}r_\delta L^{[\alpha,\beta',\gamma',\delta]}\over(\llbar{\alpha})^2(\llbar{\ubbm\alpha})^2}\right).
\end{equation*}
So $b=b^{(c)}+Q\Lambda^{(c)}$ in the patch where $\llbar{\alpha}\neq 0$. The derivation of $b=b^{(a)}+Q\Lambda^{(a)}$ in the patch where $\llbar{\dot{\alpha}}\neq 0$ is completely analogous.

\section{Twisted Field Zero Modes}\label{app:twistedzm}

This section explains how the number of zero modes of twisted fields are obtained from the general Riemann-Roch theorem given in \cite{abmnv}. Fermionic or bosonic $(1,0)$ systems are worldsheet field theories whose action is of the form
\begin{equation}
S=\int_{\Sigma_g}(b\bar{\partial}c+\tilde{b}\partial\tilde{c})
\end{equation}
where $\bar{\partial}$ is the Cauchy-Riemann operator acting on some space of objects appropriately defined, $b$ is a section of some holomorphic line bundle
$\xi$, and $c$ is a section of $K\otimes\xi^{-1}$ where $K$ is the holomorphic cotangent bundle of the Riemann surface so that the integrand is a $(1,1)$-form which can be integrated over $\Sigma_g$. 
For untwisted fields, it is well known that the number of zero modes of $b$ and $c$ are related by the Riemann-Roch theorem,
\begin{equation}\label{Riemann-Roch}
	n(\xi)-n(K\otimes\xi^{-1})={\rm deg}\!\ \xi+1-g,
\end{equation}
where ${\rm deg}\!\ \xi$ denotes the degree of the line bundle $\xi$.

In the familiar cases of fermionic systems $(b,c)$ of respective conformal weights $(\lambda,1-\lambda)$ (where $\lambda$ is taken to be an integer for simplicity), the line bundle $\xi$ is given by ${\mathcal{L}}_b\equiv K^{\otimes\lambda}$. The degree of $K$ is equal to $2g-2$, so ${\rm deg}\!\ {\mathcal{L}}_b=2\lambda(g-1)$. Plugging into (\ref{Riemann-Roch}) one gets the well known relation
\begin{equation}
n(b)-n(c)=(2\lambda-1)(g-1).
\end{equation}
In the case of $\lambda=1$, ${\mathcal{L}}_b=K$ and $c$ is just a section of the trivial line bundle over $\Sigma_g$. Thus, $c$ has $1$ zero mode (the constant function) and $b$ has $g$ zero modes which can be specified by the basis $\{\omega_i,\ i=1,...,g\}$ of holomorphic $1$-differentials on the Riemann surface.

On the other hand, for worldsheet fields $(b,c)$ of conformal weights $(1,0)$ but which obey non-trivial boundary conditions along the cycles $a_i$, $b_i$ of $\Sigma_g$, $b$ is a section of a line bundle $\xi_h$ of degree $2g-2$ which is not equivalent to $K$ and where the index $h$ parametrizes inequivalent isomorphism classes of these line bundles. $\xi_h$ is actually of the form $K\otimes\psi_h$, where $\psi_h$ is a non-trivial line bundle of zero degree. This implies that $c$ has no zero modes, $n(c)=0$, since twisted line-bundles of zero degree do not admit single-valued holomorphic sections (the trivial line bundle is the only one which does), and that
$b$ has $g-1$ zero modes which are given by a basis of the so-called $h$-twisted one-differentials, $\omega_{h,i}$, $i=1,...,g-1$ \cite{agnt,bernardtwisted}.

The question of which inequivalent twisted line bundle is represented by which twisted $(1,0)$ system is answered by noticing that the parameter $h$ is specified by the phases appearing in the twisted boundary conditions $\{h^{(a_j)},h^{(b_j)},\ j=1,...,g\}$. To see this, use the fact that holomorphic line bundles can be characterized via holonomy: given a line bundle $\xi$, if one parallel-transports an element $v$ of the fiber at some point $P\in\Sigma_g$ around a closed loop ${\mathcal{C}}$, one reaches another element $v'$ of the same fiber which is just ${\mathbb{C}}$, so that they are related by 
\begin{equation}
v'=ve^{2\pi iH({\mathcal{C}},\xi)}.
\end{equation}
The holonomy $H({\mathcal{C}},\xi)$ turns out to be real, and its exponential in the previous equation is well-defined under change of trivialization. Moreover, the holonomy changes under a deformation of ${\mathcal{C}}$ to a nearby curve ${\mathcal{C}}'$ by a quantity proportional to the integral of the curvature $R$ over the surface whose boundary is ${\mathcal{C}}'-{\mathcal{C}}$. Thus, for a flat line bundle $\psi_h$, $H$ depends only on the homology class of ${\mathcal{C}}$. This means that the holonomy defines a real cohomology class, modulo an integral class, $H(\psi)\in H^1(\Sigma;{\mathbb{R}})/H^1(\Sigma;{\mathbb{Z}})$; therefore, flat line bundles are characterized by phases $h\equiv\{h^{(a_i)},\ h^{(b_i)},\ i=1,...,g\}$ around the non-trivial homology cycles of $\Sigma_g$. This establishes the connection with $(b,c)$ systems with non-trivial boundary conditions around cycles of $\Sigma_g$.

\section{Holomorphic Factorization of the Twisted Partition Function}\label{app:twistedpf}

In this section, the holomorphic structure of the twisted chiral partition function is derived using the bosonization formula of \cite{abmnv}. 



For a pair of fermions $(b,c)$ of conformal weight $(1,0)$ in the twisted sector where $c$ has no zero modes, the relevant bosonization formula is
\begin{equation}\label{fb}
\left\langle\prod_{i=1}^{g-1}\big\|b(P_i)\big\|^2\right\rangle=\left\langle \prod^{g-1}_{i=1} e^{i\varphi(P_i)}\right\rangle_{boson}
\end{equation}
for an appropriately chosen field theory of a compact boson $\varphi$  \cite{abmnv}. In this formula, the norm is defined by $\|b(P)\|^2=\rho^{-1}(P)b(P)\tilde{b}(P)$ where $\rho(P)$ defines the conformally flat metric $ds^2=\rho dzd\bar{z}$. 
%
Computing the boson correlator in the right hand side of (\ref{fb}), one obtains 
\begin{equation}\label{bf1}
\left\langle\prod_{i=1}^{g-1}\big\|b(P_i)\big\|^2\right\rangle =
|Z_1|^{-1} {\mathcal{N}}(z_\xi)\prod^{g-1}_{i<j}G(P_i,P_j)^2,
\end{equation}
where 
$|Z_1|^{-1}$ is the partition function of a free scalar field, 
\begin{equation}
{\mathcal{N}}(z_\xi)=e^{-2\pi{\rm Im}z_\xi({\rm Im}\tau)^{-1}{\rm Im}z_\xi}|\Theta(z_\xi | \tau)|^2,
\end{equation}
$z_\xi$ is defined for a twisted line bundle with phases $h^{(a)}$ and $h^{(b)}$ around the $a$ and $b$ cycles of $\Sigma_g$ as \cite{abmnv}
\begin{equation}
	z_{{\xi}}=h^{(b)}+\tau h^{(a)}-\sum^{g-1}_{i=1}P_i+\Delta,
\end{equation}
and $G(P_i,P_j)$ is the Green's function of the Laplacian which depends on the specific metric chosen on the Riemann surface $\Sigma_g$. There is a special metric \cite{faykernel} that simplifies computations considerably:
\begin{equation}
\rho(P)=|\sigma(P)|^{4/g}\exp\left[{4\pi\over g(g-1)}{\rm Im}\Delta^P({\rm Im}\tau)^{-1}{\rm Im}\Delta^P\right],
\end{equation}
where $\Delta^P$ is defined as $\displaystyle\Delta^P=\int^\Delta_{(g-1)P}\omega$. With this choice, one gets
\begin{equation}
G(P_i,P_j)^2=\rho^{1/2}(P_i)\rho^{1/2}(P_j)F(P_i,P_j),
\end{equation}
where $F(P_i,P_j)$ is defined as
\begin{equation}
F(P_i,P_j)=\exp\left[-2\pi{\rm Im}\!\int^{P_i}_{P_j}\!\omega\ ({\rm Im}\tau)^{-1}{\rm Im}\!\int^{P_i}_{P_j}\!\omega\right]|E(P_i,P_j)|^2
\end{equation}
and $E(P_i,P_j)$ is the so-called prime form.

To express (\ref{fb}) as a holomorphic square, it is convenient to multiply and divide the left-hand side of (\ref{bf1}) by the formula for the untwisted partition function \cite{orbinsv}.
In the untwisted sector, $c$ has a zero mode and the relevant bosonization formula is 
\begin{equation}\label{bf2}
\left\langle\prod_{i'=1}^g\big\|b(P_{i'}) \big\|^2 | c(Q)|^2 \right\rangle 
= 
|Z_1|^{-1} {\mathcal{N}}(z_K){\prod^{g}_{i'<j'}G(P_{i'},P_{j'})^2\over\prod^{g}_{i'=1}G(P_{i'},Q)^2}
\end{equation}
where $z_{K}=-\sum_{i'=1}^g P_{i'}+Q+\Delta$. Since 
\begin{equation}\label{bff2}
\left\langle\prod_{i'=1}^g\big\|b(P_{i'}) \big\|^2 | c(Q)|^2 \right\rangle
= |Z_1|^2 \|{\rm det}(\omega_{i'}(P_{j'}))\|^2,
\end{equation}
one can multiply and divide the left-hand side of (\ref{bf1}) by this partition function to
obtain the formula

\begin{equation}
\left\langle\prod_{i=1}^{g-1}\big|b(P_i)\big|^2\right\rangle
 = \rho^{-1}(P_g)|{\rm det}\omega_{i'}(P_{j'})|^2|Z_1|^2\ {{\mathcal{N}}\big(z_{\xi}\big)\over{{\mathcal{N}}\big(z_{{K}}\big)}}\ {\prod^{g}_{i'=1}G(P_{i'},Q)^2\over\prod^{g-1}_{i=1}G(P_i,P_g)^2}
\end{equation}
where one chooses $P_{i'}=P_i$ for $i=i'=1,...,g-1$ and $P_{i'=g}\equiv P_g$ and
$Q$ are arbitrary points.
Taking the limit as $Q\rightarrow P_g$, it is straightforward to show that the result is a holomorphic square
\begin{equation}
\left\langle\prod_{i=1}^{g-1}\big|b(P_i)\big|^2\right\rangle
=|{\rm det}(\omega_{i'},P_{j'})|^2|Z_1|^2{\left|\Theta{\left[\hspace{-0.1cm}\begin{array}{c} h^{(a)} \\ \vspace{-0.45cm} \\ h^{(b)} \end{array}\hspace{-0.1cm}\right]}(z|\tau)\right|^2\over\displaystyle\left|\sum_{i'=1}^g\partial_{i'}\Theta(z|\tau)\omega_{i'}(P_g)\right|^2},
\end{equation}
where $z=\Delta-\sum^{g-1}_{i=1}P_i$, and the theta function with characteristics in the numerator is defined as
\begin{equation}
\Theta{\left[\hspace{-0.1cm}\begin{array}{c} \varphi \\ \vspace{-0.55cm} \\ \phi \end{array}\hspace{-0.1cm}\right]}(z|\tau)=e^{i\pi\varphi\cdot(\tau\cdot\varphi+2(z+\phi))}\Theta(\phi+\tau\cdot\varphi+z|\tau).
\end{equation}

So finally one can define a twisted chiral partition function $Z_{1,h}$ satisfying
\begin{equation}
 \left\langle\prod_{i=1}^{g-1}\big|b(P_i)\big|^2\right\rangle =
 |Z_{1,h} det (\omega_{h,i}(P_j))|^2
 \end{equation}
 by the formula
\begin{equation}
Z_{1,h}{\rm det}(\omega_{h,i}(P_j))=Z_1{\rm det}(\omega_{i'},P_{j'}){\Theta{\left[\hspace{-0.1cm}\begin{array}{c} h^{(a)} \\ \vspace{-0.45cm} \\ h^{(b)} \end{array}\hspace{-0.1cm}\right]}(z|\tau)\over\displaystyle\sum_{i'=1}^g\partial_{i'}\Theta(z|\tau)\omega_{i'}(P_g)},
\end{equation}
where the choice of $P_g$ is arbitrary.

\end{document}